\RequirePackage{fix-cm}
\documentclass[final,times,3p,twocolumn,epjc3]{svjour3}  
\smartqed  % flush right qed marks, e.g. at end of proof
\RequirePackage{graphicx}
\RequirePackage{amsmath}
\RequirePackage{lineno,hyperref}
\RequirePackage{upgreek}
\RequirePackage{xcolor}
\RequirePackage{siunitx}
\journalname{Eur. Phys. J. C}
\begin{document}

%\title{The Extreme Energy Events (EEE) telescope simulation framework}%\thanksref{t1}
 \title{The cosmic muon and detector simulation framework of the Extreme Energy Events (EEE) experiment} % Titolo proposto da Marcello...controllare

%\titlerunning{EEE simulation framework}       % if too long for running head

\author{M.~Abbrescia\thanksref{BariUniv,BariINFN} \and
C.~Avanzini\thanksref{PisaINFN,CentroFermi}\and 
L.~Baldini\thanksref{PisaINFN,CentroFermi,PisaUniv} \and 
R.~Baldini~Ferroli\thanksref{CentroFermi,FrascatiINFN} \and
G.~Batignani\thanksref{PisaINFN,CentroFermi,PisaUniv} \and 
M.~Battaglieri\thanksref{GenovaINFN,JLab,Corresponding} \and 
S.~Boi\thanksref{CentroFermi,CagliariUniv,CagliariINFN} \and
E.~Bossini\thanksref{CERN} \and 
F.~Carnesecchi\thanksref{CentroFermi,BolognaINFN,BolognaUniv} \and 
C.~Cical\`{o}\thanksref{CentroFermi,CagliariINFN} \and
L.~Cifarelli\thanksref{CentroFermi,BolognaINFN,BolognaUniv} \and 
F.~Coccetti\thanksref{CentroFermi} \and 
E.~Coccia\thanksref{CentroFermi,GSSI} \and
A.~Corvaglia\thanksref{CentroFermi,LecceINFN} \and 
D.~De~Gruttola\thanksref{SalernoUniv,SalernoINFN} \and 
S.~De~Pasquale\thanksref{SalernoUniv,SalernoINFN} \and
F.~Fabbri\thanksref{CentroFermi,FrascatiINFN} \and 
A.~Fulci\thanksref{MessinaUniv} \and
L.~Galante\thanksref{CentroFermi,TorinoPolitScienze,TorinoINFN} \and 
M.~Garbini\thanksref{CentroFermi,BolognaINFN} \and 
G.~Gemme\thanksref{GenovaINFN} \and 
I.~Gnesi\thanksref{CentroFermi,CosenzaINFN} \and
S.~Grazzi\thanksref{CentroFermi,GenovaINFN,Corresponding} \and 
D.~Hatzifotiadou\thanksref{BolognaINFN,CERN,CentroFermi} \and 
P.~La~Rocca\thanksref{CentroFermi,CataniaUniv,CataniaINFN} \and 
Z.~Liu\thanksref{WorldLaboratory} \and
G.~Mandaglio\thanksref{CentroFermi,CataniaINFN,MessinaUniv,Corresponding} \and
G.~Maron\thanksref{BolognaCNAF} \and 
M.~N.~Mazziotta\thanksref{BariINFN} \and
A.~Mulliri\thanksref{CagliariUniv,CagliariINFN} \and 
R.~Nania\thanksref{BolognaINFN,CentroFermi} \and 
F.~Noferini\thanksref{CentroFermi,BolognaINFN} \and
F.~Nozzoli\thanksref{TrentoINFN} \and
F.~Palmonari\thanksref{CentroFermi,BolognaINFN,BolognaUniv} \and
M.~Panareo\thanksref{LecceINFN,LecceUniv} \and
M.~P.~Panetta\thanksref{CentroFermi,LecceINFN} \and 
R.~Paoletti\thanksref{PisaINFN,SienaUniv} \and 
C.~Pellegrino\thanksref{CentroFermi,BolognaCNAF} \and
%L.~Perasso\thanksref{GenovaINFN,CentroFermi} \and
O.~Pinazza\thanksref{BolognaINFN,CentroFermi} \and
C.~Pinto\thanksref{CataniaUniv,CataniaINFN,CentroFermi} \and
S.~Pisano\thanksref{FrascatiINFN,CentroFermi} \and
F.~Riggi\thanksref{CentroFermi,CataniaUniv,CataniaINFN} \and 
G.~Righini\thanksref{CNR} \and
C.~Ripoli\thanksref{SalernoUniv,SalernoINFN} \and 
M.~Rizzi\thanksref{BariINFN} \and 
G.~Sartorelli\thanksref{CentroFermi,BolognaINFN,BolognaUniv} \and
E.~Scapparone\thanksref{CentroFermi,BolognaINFN} \and 
M.~Schioppa\thanksref{CosenzaINFN,CosenzaUNIV} \and 
A.~Scribano\thanksref{SienaUniv} \and
M.~Selvi\thanksref{CentroFermi,BolognaINFN} \and 
G.~Serri\thanksref{CentroFermi,CagliariUniv,CagliariINFN}
S~ Squarcia\thanksref{GenovaINFN,GenovaUniv} \and
M.~Taiuti\thanksref{GenovaINFN,GenovaUniv} \and 
G.~Terreni\thanksref{PisaINFN,CentroFermi} \and 
A.~Trifir\`{o}\thanksref{CentroFermi,CataniaINFN,MessinaUniv} \and 
M.~Trimarchi\thanksref{CentroFermi,CataniaINFN,MessinaUniv} \and 
A.S.~Triolo\thanksref{MessinaUniv} \and
C.~Vistoli\thanksref{BolognaCNAF} \and 
L.~Votano\thanksref{GranSassoINFN} \and 
M.~Ungaro\thanksref{JLab} \and
M.~C.~S.~Williams\thanksref{CentroFermi} \and 
A.~Zichichi\thanksref{CentroFermi,BolognaINFN,CERN,BolognaUniv} \and 
R.~Zuyeuski\thanksref{CentroFermi}        
}

\institute{Dipartimento Interateneo di Fisica, Universit\`{a} di Bari, Bari, Italy  \label{BariUniv} \and
INFN Sezione di Bari, Bari, Italy  \label{BariINFN} \and
INFN Sezione di Pisa, Pisa, Italy \label{PisaINFN} \and
Museo storico della Fisica e Centro studi e ricerche ``E. Fermi", Roma, Italy  \label{CentroFermi} \and
INFN Sezione di Bologna, Bologna, Italy \label{BolognaINFN} \and
Dipartimento di Fisica, Universit\`{a} di Pisa, Pisa, Italy \label{PisaUniv} \and
INFN, Laboratori Nazionali di Frascati, Frascati (RM), Italy \label{FrascatiINFN} \and
INFN Sezione di Genova, Genova, Italy \label{GenovaINFN} \and
Thomas Jefferson National Accelerator Facility, Newport News, VA 23606, USA \label{JLab} \and
Dipartimento di Fisica, Universit\`{a} di Cagliari, Cagliari, Italy \label{CagliariUniv} \and
INFN Sezione di Cagliari, Cagliari, Italy \label{CagliariINFN}\and
CERN, Geneva, Switzerland \label{CERN} \and
Dipartimento di Fisica ed Astronomia, Universit\`{a} di Bologna, Bologna, Italy \label{BolognaUniv} \and
Gran Sasso Science Institute, Italy \label{GSSI} \and
INFN Sezione di Lecce, Lecce, Italy \label{LecceINFN} \and
Dipartimento di Fisica, Universit\`{a} di Salerno, Salerno, Italy \label{SalernoUniv} \and
INFN Gruppo Collegato di Salerno, Salerno, Italy \label{SalernoINFN} \and
Dipartimento di Scienze Matematiche e Informatiche, Scienze Fisiche e Scienze della Terra, Universit\`{a} di Messina, Messina, Italy \label{MessinaUniv} \and
Dipartimento di Scienze Applicate e Tecnologia, Politecnico di Torino, Torino, Italy 
\label{TorinoPolitScienze} \and
INFN Sezione di Torino, Torino, Italy \label{TorinoINFN} \and
INFN Gruppo Collegato di Cosenza, Laboratori Nazionali di Frascati (RM), Italy \label{CosenzaINFN} \and
Dipartimento di Fisica e Astronomia, Universit\`{a} di Catania, Catania, Italy \label{CataniaUniv} \and
INFN Sezione di Catania, Catania, Italy \label{CataniaINFN} \and
ICSC World laboratory, Geneva, Switzerland \label{WorldLaboratory} \and
INFN-CNAF, Bologna, Italy \label{BolognaCNAF} \and
INFN Trento Institute for Fundamental Physics and Applications, Trento, Italy \label{TrentoINFN} \and 
Dipartimento di Matematica e Fisica, Universit\`{a} del Salento, Lecce, Italy \label{LecceUniv} \and
Dipartimento di Scienze Fisiche, della Terra e dell’Ambiente, Universit\`{a} di Siena, Siena, Italy \label{SienaUniv} \and
CNR, Istituto di Fisica Applicata ``Nello Carrara", Sesto Fiorentino, Italy \label{CNR} \and
Dipartimento di Fisica, Universit\`{a} della Calabria, Rende (CS), Italy \label{CosenzaUNIV} \and
Dipartimento di Fisica, Universit\`{a} di Genova, Genova, Italy \label{GenovaUniv} \and
INFN, Laboratori Nazionali del Gran Sasso, Assergi (AQ), Italy \label{GranSassoINFN}
}

\thankstext[$^\star$]{Corresponding}{Corresponding authors:\\ Marco Battaglieri, e-mail: marco.battaglieri@ge.infn.it,\\ Stefano Grazzi, e-mail: stefano.grazzi@ge.infn.it,\\ Giuseppe Mandaglio, e-mail: gmandaglio@unime.it}

\date{Received: date / Accepted: date}

\maketitle
\begin{abstract}
    
This paper describes the simulation framework of the Extreme Energy Events (EEE) experiment. EEE is a network of cosmic muon trackers, each made of three Multi-gap Resistive Plate Chambers (MRPC), able to precisely measure the absolute muon crossing time and the muon integrated angular flux at the ground level. The response of a single MRPC and the combination of three chambers have been implemented in a GEANT4-based framework (GEMC) to study the telescope response. The detector geometry, as well as details about the surrounding materials and the location of the telescopes have been included in the simulations in order to realistically reproduce the experimental set-up of each telescope.
A model based on the latest parametrization of the cosmic muon flux has been used to generate single muon events. After 
validating the framework by comparing simulations to selected EEE telescope data, it has been used to determine detector parameters not accessible by analysing experimental data only, such as detection efficiency, angular and spatial resolution. 
\end{abstract}

\section{Introduction}
\label{intro}
The Extreme Energy Events (EEE) experiment~\cite{ee1,ee2} has deployed a network of about 60 cosmic muon detectors sparse in an area of 3 $\times$ 10$^{5}$ {km}$^{2}$.\\ 
The EEE network acts as a gigantic telescope that, precisely measuring cosmic muon rates and arrival times, looks at the sky in a complementary way than traditional optical telescopes. The EEE main goal is to study high-energy cosmic rays. Some recent results, published by the EEE Collaboration, include: observation of the Forbush effect~\cite{eean1}, searches for anisotropies in the cosmic ray intensity~\cite{eean2}, and long distance correlation in secondary muons~\cite{ee2}.\\
The study of the local cosmic muon flux and its dependence on environmental and astrophysical conditions provides insight into the Forbush effect, a temporary reduction in the galactic cosmic rays flux subsequent to Solar flares accompanied by Coronal Mass Ejections \footnote{The effect is explained considering the magnetic field of the plasma solar wind sweeping part of the galactic cosmic rays away from the Earth.}.
Anisotropies in the cosmic ray intensity induced by large scale or local magnetic field features, or by the Compton-Getting effect, have been reported but more extensive observations are needed for a quantitative understanding of this effect. Finally, the search for long distance correlations in secondary muons produced by extreme energy events, distributed over distances up to 1200 km and detected by different detectors of the EEE network is a hot topic, since it could point out a complete new feature in cosmic ray physics and be of striking importance for multi-messenger astronomy.\\
A quantitative understanding of these effects can be obtained by a precise measurement of the cosmic muon rates and arrival times (at ns level). To do so a sophisticated detector, a good understanding of the experimental conditions and an excellent synchronization of data coming from different telescopes in the EEE grid is mandatory.
All EEE detectors are based on the same MRPC technology, (described in the next Section) but their experimental configurations are slightly different (e.g. the distance between the chambers and the absolute orientation w.r.t. the North are not always the same). Moreover,
the measured rate is affected by the material surrounding the detector that is different for each telescope since they are hosted in rooms located in non-dedicated buildings (high schools or university labs). For a thorough interpretation of experimental observations (cosmic ray absolute rates and angular distributions), and, following a standard methodology in particle physics, a Monte Carlo simulation of the detectors response and experimental conditions is therefore necessary.\\
In this article we describe the EEE simulation framework (based on GEANT4 libraries~\cite{geant}) and its validation. It includes: single cosmic muon generation, propagation through materials surrounding the detector and a parametric description of the MRPC response to charged particles. This is a first step for a comprehensive simulation of the EEE network response to cosmic muon showers that includes: generation of a high energy primary cosmic ray, interaction with the high atmosphere, production of a cascade of secondary particles and propagation down to the ground trough the air and other materials surrounding the EEE detectors. For this purpose, the simulation framework is being currently interfaced to the CORSIKA cosmic shower generator~\cite{corsika}. A separate publication with results of this study is in preparation.\\
The paper is organized as follows: the working principle of an EEE
muon telescope is described in 
Sec.~\ref{eee_tel} and its implementation in the simulation framework in Sec.~\ref{sim_frame}; the validation of the framework and the comparison of simulations with data are reported in Sec.~\ref{sim_frame_valid}; finally, systematic studies on detector performance in various experimental set-ups are reported in Sec.~\ref{eee_tel_sim}.

%The paper is organized as follows: in Sec.\ref{sim_frame} the EEE simulation framework is described; the validation of the framework and the comparison of simulations with data are reported in Sec.~\ref{sim_frame_valid}; finally, systematic studies on detector performance and experimental set-ups are reported in Sec.~\ref{eee_tel_sim}.

\section{The EEE Telescope}
\label{eee_tel}

%\subsection{The EEE Telescope}

In this section we briefly describe the EEE telescope features useful to understand why the simulation framework was structured in the way described later on. More details can be found in Ref.~\cite{ee1}.
\begin{figure}[th!]
%\centering 
\hspace{-0.8cm}
\includegraphics[width=1.05\columnwidth]{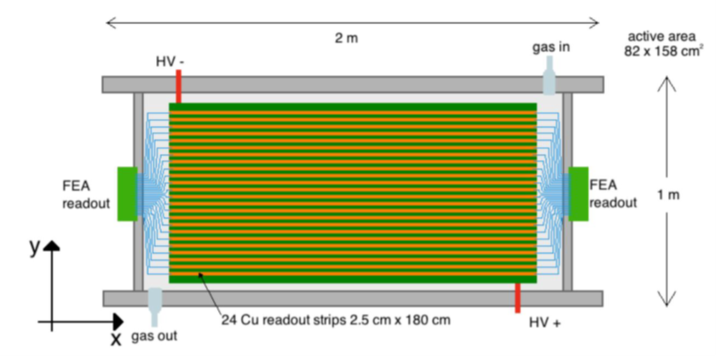} \caption{Top view of an EEE MRPC chamber.}
\label{mrpc_top}
\end{figure}

\begin{figure}[th!]
\centering 
\includegraphics[width=.8\columnwidth]{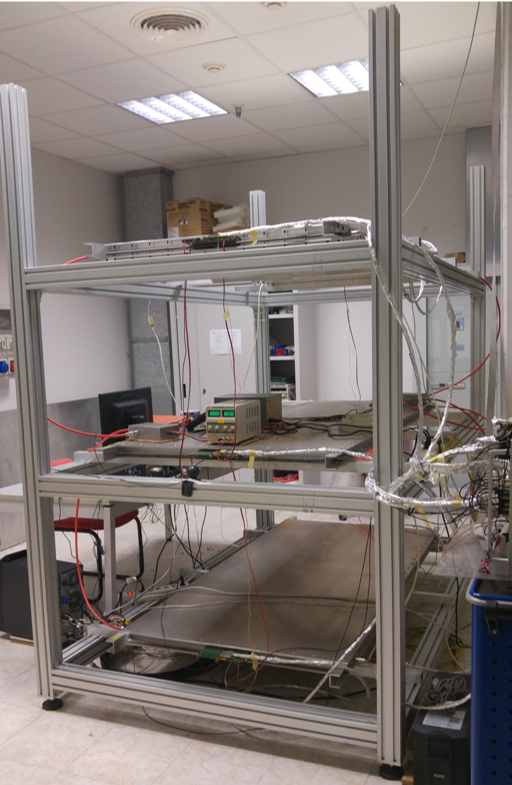} \caption{Photograph of an EEE telescope.}
\label{eee_telescope} 
\end{figure}

Each station of the EEE network, that defines a ``telescope" for cosmic rays (mainly muons), is made of three Multigap Resistive Plate Chambers (MRPC)~\cite{mrpc} specifically designed to achieve good tracking and timing capability, low construction costs, and an easy assembly procedure~\cite {mrpc1}. A schematic of an MRPC is shown in Fig~\ref{mrpc_top}.
The three MRPC chambers are placed one above the other with the top and the bottom chambers at a distance of 50~cm from the middle chamber (referred, in the following, as 50/50~cm or ``standard'' configuration) resulting in an angular acceptance of 2.23 sr. 
%(minor review)\footnote{Some telescopes have the three MRPCs located at a different distance: for example CERN-01, 44/44~cm, and SAVO-01, 46/46~cm, and therefore provide a different angular acceptance.}
A photograph of an EEE telescope is shown in Fig.~\ref{eee_telescope}.

%Each MRPC consists of 6 gas gaps obtained by stacking glass sheets separated from each other by fishing line spacers, placed inside a gas-tight aluminum box.
%(stefano) The active area of MRPC (158~cm $\times$ 82~cm) consists of 6 gas-gaps obtained by stacking glass sheets, separated from each other by fishing line spacers, with voltage applied only to the first and the last, and leaving the inner ones floating. 
%%On the outer surfaces a sheet of Mylar is stretched on a vetronite panel of equal area on whose external surface 24 copper strips are laid out , to collect the signals induced by particles. 
%A set of 24 copper strips (Fig~\ref{mrpc_top}) are mounted on both sides of the stack of glass plates to collect the signals induced by avalanches generated by particles into the gas, forming a differential signal between cathode and anode strips. Two Front-End Amplifier Boards (FEA) placed at the two sides records the strips signal. Finally, Two rigid larger composite honeycomb panels are used to assure good mechanical stability to the whole structure, enclosed in a gas-tight aluminum box. 
%(Stefano)Chambers are filled with a gas mixture consisting of a 98/2 combination of R134a (C2F4H2) and SF6, at a continuous flow of 2 l/h and atmospheric pressure. The high voltage (HV) to generate the multiplication field is provided by a set of DC/DC converters. The total HV applied on the chambers is in the 18 to 20 kV range.

Each MRPC consists of 6 gas gaps obtained by stacking glass sheets separated from each other by fishing line spacers, placed inside a gas-tight aluminum box and flushed with a mixture made of C$_2$H$_2$F$_4$ and SF$_4$ in 98/2 proportions. 
%The active area of each chamber is 82~cm $\times$ 158~cm, and t
The readout panel is split in 24 copper strips, 180~cm long and 2.5~cm wide each, separated one from another by 7 mm of space, each connected to a readout channel of the EEE Front-End Amplifier Boards (FEA). The active area is defined by the 158 cm x 82 cm glass surface.
High voltage electrodes placed on the opposite sides of the stack provide an electric field in the gas, strong enough that the ion electron pairs produced by ionizing particles crossing the MRPCs give origin to Townsend avalanches which induce detectable signals on the readout strips. The high voltage (HV) to generate the multiplication field applied on the chambers is in the 18 to 20 kV range.\\

\subsection{Detector working principle}
%The active area of each chamber is 158~cm $\times$ 82~cm.
%and the readout panel is split in 24 readout channels in the form of copper strips, 158 $\times$ 2.5~cm each. Signals are processed (amplified and discriminated) by the NINO ASIC~\cite{nino-asic}, and digitized by a CAEN-v1190 multi-hit TDC (Time-To-Digit converter). The muon impact point X-coordinate is given by the arrival time difference of the signal at the two edges of the readout strip, while the Y-coordinate is obtained by the channel position.
In the single MRPC, signals from copper strips are processed (amplified and discriminated) by the NINO ASIC~\cite{nino-asic}, and digitized by a CAEN-v1190 multi-hit TDC (Time-To-Digital converter). Each strip corresponds to a specific channel of  the TDC system. \\
%(Stefano)The readout electronics used in the EEE project is such that the measured hit time is given by the time when the signal on each strip becomes larger than the threshold. 
A muon track candidate is identified by requiring that, at least one of the 24 channels of each of the six FEA (one for each short sides of the three chambers of a telescope) has a valid signal.
When this condition is achieved a trigger signal is generated by the EEE Trigger Board~\cite{EEETriggerBoard}, and used as a start to the TDCs. 
The track's impact point (hit) X-coordinate is derived  by the arrival time difference of the signal at the ends of the readout strip (measured by TDC). The hit Y coordinate is simply obtained by the corresponding strip position or from the middle position in case of crossing between two adjacent strips.\\
To improve synchronization between MRPCs, a common clock is generated by a custom VME board and distributed to the TDCs. The time resolution at the single chamber level, given by the convolution of the intrinsic MRPC time resolution and the uncertainties introduced by the signal processing chain, is about 250 ps. Synchronization between telescopes, sometimes located hundreds of km away, is guaranteed by a GPS unit that provides an absolute time stamp with a precision at the level of 40 ns~\cite{gps}.

 The EEE MRPC telescopes show good performance in terms of efficiency of tracks reconstruction\footnote{The reconstruction efficiency is defined as the percentage of raw events where at least one track candidate has been identified.}($\epsilon \approx 93\%$), time resolution ($\sigma_T \approx 238$ ps), and spatial resolution ($\sigma_{S_L} \approx 1.5$~cm and $\sigma_{S_T} \approx 0.92$~cm for longitudinal and transverse directions, respectively)~\cite{performance}. 

%(minor review)\subsection {Reference telescopes}
%\label{tori03}

The telescopes are located inside high schools buildings, universities and research centres, operated by students and teachers under the supervision of EEE Collaboration researchers (no telescopes are installed outdoor).
Dense materials, such as concrete roof or walls, or iron, surrounding the telescopes, shield the detector from low energy charged particles (e.g. electrons) leaving only cosmic muons as track candidates. Low energy electrons deriving from interactions of muons in the bottom layers of concrete have a low probability to cross the three MRPCs and give rise to a trigger.

The EEE telescopes operate continuously from September to June following the school calendar.
From 2014 a total of more than 100 billions of muon track candidates have been collected.
Data are transferred daily from the telescope stations to the INFN-CNAF data center in Bologna for storage and further analysis~\cite{cnaf}.

\subsection {Reference telescopes}
\label{tori03}
\begin{figure}[th!]
\flushleft
\includegraphics[width=0.9\columnwidth]{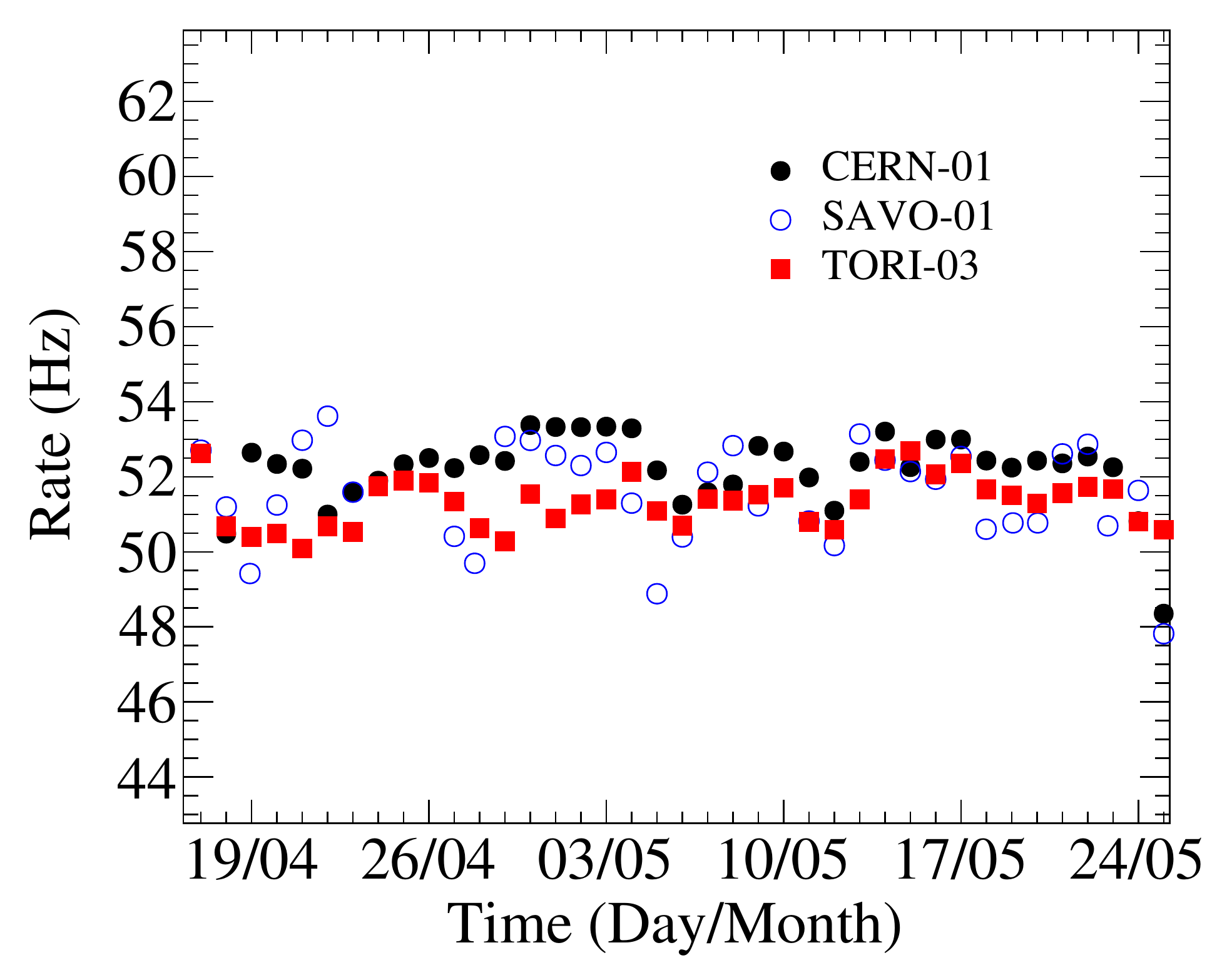}
\caption{Counting rate for the EEE stations CERN-01, SAVO-01 and TORI-03, from April 17, 2018 up to May 25, 2018. Error bars are inside the marker.} 
\label{fig:daq_rate_run3}
\end{figure}

Among the $\approx$60 telescopes in operation within the EEE network, we selected as a reference to validate the framework some telescopes know to be very stable in time  and hosted in building with roof and walls easy-to-implement in simulations. The telescope TORI-03,
hosted in the High School ``Liceo Scientifico Statale Galileo Ferraris" in Torino and active since 2014, is located in a room at the uppermost floor of the school.
The CERN-01 telescope, hosted by CERN, is located in a building with thin walls and roof with top/bottom chambers distanced by 44/44~cm. Finally, the SAVO-01 telescope, is located in a room at the uppermost floor of ``Liceo Statale Chiabrera-Martini" in Savona with the three chambers 46/46~cm apart.
The single muon counting rate for TORI-03, CERN-01, and SAVO-01 are shown in Fig.~\ref{fig:daq_rate_run3}.

\section{Simulation Framework}
\label{sim_frame}

A detailed description of the MRPC geometry and materials has been implemented in GEMC~\cite{gemc}, a GEANT4 libraries-based interface with user-defined geometry and hits description~\cite{geant}\footnote{For these studies we used version 4.10.03.p02 of GEANT libraries: $FTFP\_BERT$ physics model for hadronic interactions \cite{bertini} to reproduce the cosmic muon crossing in telescope surrounding materials and 
Wentzel VI Model~\cite{went} for an accurate descriprion of 
electromagnetic processes (in particular multiple and large angle scattering).}. 
\begin{figure}[th!]
\centering 
\includegraphics[width=.9\columnwidth]{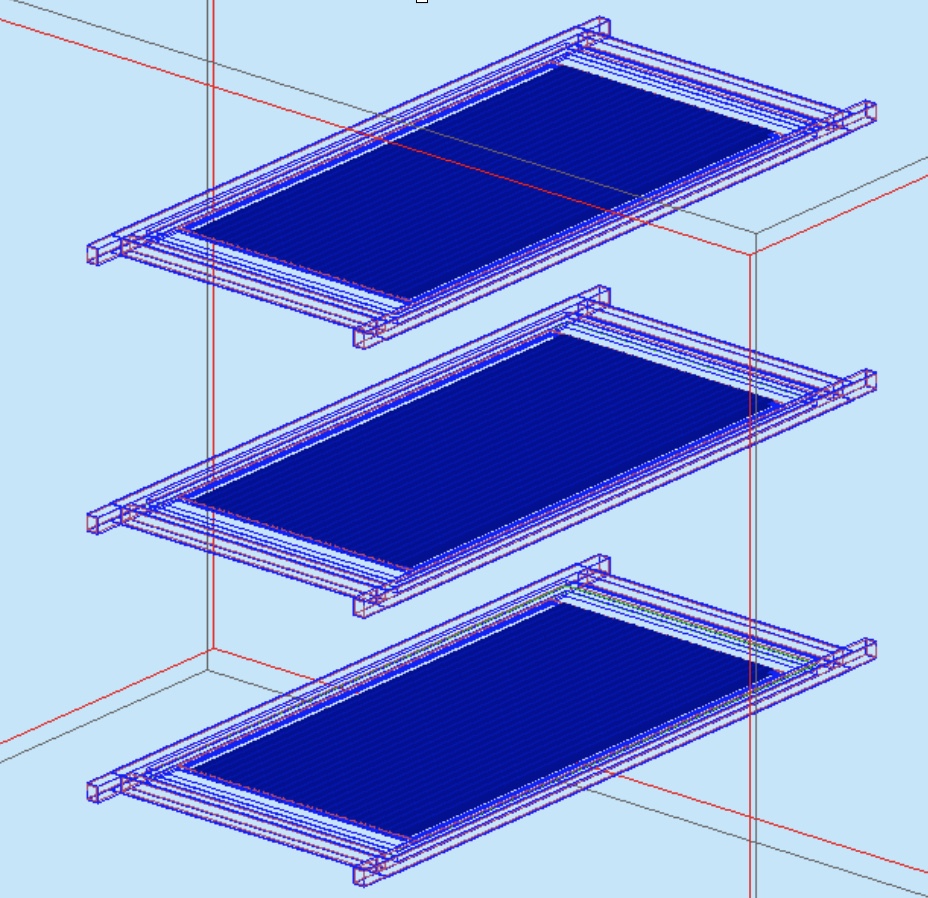} 
\caption{The three MRPCs forming an EEE telescope, as implemented in GEMC.}
\label{fig:MRPC-telescope} 
\end{figure}
 The external frame, made from 220 cm x 92 cm x 0.4 cm Al plates, contains the detector components as follow (starting from the bottom to the top): the Al bottom honeycomb panel (180 cm x 90 cm x 1.5 cm), the bottom vetronite panel (180 cm x 90 cm x 0.15 cm), the mylar sheet (180 cm x 90 cm x 0.0175 cm), the bottom large glass panel (160 cm x 85 cm x 0.2 cm), six smaller glass panels (160 cm x 85 cm x 0.15 cm), separated by an gap of 0.025 mm (corresponding to the size of the fishing line used as spacer), and again, the top large glass, mylar sheet, vetronite and Al honeycomb panels of the same sizes as of bottom panels. The readout pads (GEANT4 active volume) have been implemented as copper strips (180 cm x 2.5 cm x 0.05 cm) spaced by 0.7 cm gaps on the top side of 
 %the top side of the bottom 
 vetronite panel. The same structure is duplicated three times with three different identifiers of the active pads.
The three chambers forming a telescope are positioned in a stack with a variable distance that reproduces the actual experimental set up. 
In the rest of the paper, we consider as "standard geometry" the configuration with the three MRPCs spaced by 50 cm from each other.
To mimic a generic room where a real telescope is located, the three chambers are inserted in a box whose walls are made of concrete with variable thickness. The {\it nominal} configuration corresponds to a thickness of 30~cm for each of the six box walls. This is the simplest configuration for an EEE telescope: more complicated geometries of the surrounding materials have also been implemented, for instance, by considering the foundations of multi-floors buildings and/or nearby obstacles, such as other buildings or natural structures. Some examples are reported in Sec.~\ref{eee_tel_sim}.
The implementation of the MRPC geometry in GEMC is shown in Fig.~\ref{fig:MRPC-telescope}.

\subsection{Detector response}
\label{detc_rep}
\begin{figure}[th!]
\centering 
\includegraphics[width=.7\columnwidth]{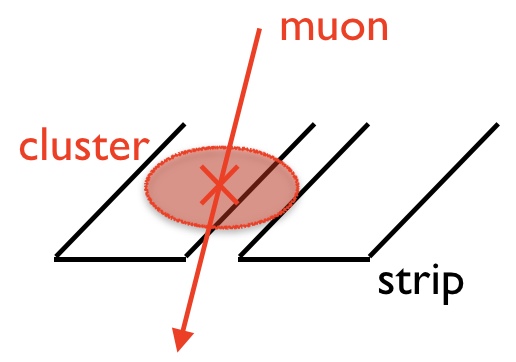} 
\caption{Effective implementation of the signal induced by an avalanche on the strip plane. The ellipse size corresponds to the measured cluster size ($\sigma_{S_L}$=0.92 cm and $\sigma_{S_T}$= 1.5 cm, see Ref.\cite{performance}).} 
\label{fig:MRPC-cluster} 
\end{figure}

The MRPC response was parametrized based upon the measured performance of the chambers. The ionization avalanche in the gas produced by a crossing charged particle was effectively described by considering its effect on the sensitive area of the detector, namely the readout plane, segmented in 2.5~cm copper strips.
The strip multiplicity, defined as the number of contiguous firing strips per particle crossing, was obtained by centering an ellipse in the intersection point of the track with the strip plane. The ellipse sizes were assumed to be the measured cluster sizes $\sigma_{S_L}$ and $\sigma_{S_T}$~\cite{performance}.  In case the ellipse covers two or more  adjacent strips, all of them are assumed to fire. This procedures takes into account multiple-hits per event recorded by the telescope.
To clarify how the induced signal has been parametrized, Fig.~\ref{fig:MRPC-cluster} shows a graphical representation of the 
strip multiplicity.
The hit time was propagated to the strip sides by considering the measured signal propagation velocity along the strip (15.8~cm/ns) with a gaussian resolution of $\sigma_T$=238 ps.
For these simulations the same set of parameters was used for every chamber but the implementation of a specific measured MRPC response is straightforward.  

\subsection {Pseudo-data reconstruction algorithm}
\label{rec_algorithm}
Events generated with GEMC are pre-processed by a program that uses part of the information to mimic the hardware trigger and generate pseudo-data to be fed to the standard EEE data reconstruction chain. 
The (pseudo) trigger is generated considering only events with at least one hit per chamber (minimal condition required by the experimental trigger).
In order to match the format used by the standard EEE data reconstruction code, the following information is included in the stream:
\begin{itemize}
    \item total number of hits;
    \item identifier (channel) of strip with signal;
    \item time interval that is necessary for avalanche generated signal to travel on strip from hit interaction point to chamber side where are the FEA;
    \item hit-time matching within a predefined trigger window. 
\end{itemize}

\subsection{Cosmic muon flux parametrization}
\label{mu_par}
Single-muon events are generated according to the model described in Ref.\cite{spheregen}.
The azimuthal $\phi_\mu$ distribution is considered to be uniform. The polar angle distribution ($\theta_\mu$ being the angle with respect to the vertical) is taken from an improved ~\cite{cosmugen} Gaisser-like parametrization~\cite{gaisser} that includes the Earth curvature at all latitudes and low energy muons ($E_\mu<100$ GeV). The flux, as a function of muon energy $E_\mu$ and $\theta_\mu$ is given by:

\begin{equation}
\begin{aligned}
        \frac{dI_\mu}{dE_\mu}=0.14
    \left[ \frac{E_\mu}{GeV} \left(1+\frac{3.64\,GeV}{E_\mu(cos(\theta^*))^{1.29}}\right) \right]^{-2.7}\\
    \times \left[ \frac{1}{1+ \frac{1.1\, E_\mu\,cos \theta^*}{115 GeV}} +
                  \frac{0.054}{1+ \frac{1.1\, E_\mu\,cos \theta^*}{850 GeV}} \right]
\end{aligned}
\end{equation}
where 
\begin{equation}
\begin{aligned}
\hspace{-0.5cm}
cos\theta^* = 0.14 \sqrt{\frac{(cos\theta_\mu)^2+ P^2_1+P_2(cos\theta_\mu)^{P_3}+P_4(cos\theta_\mu)^{P_5}}{1+P_1^2+P_2+P_4}}
\end{aligned}
\label{costheta}
\end{equation}
with parameters reported in Table \ref{tab:my_label}. The absolute muon flux is normalized to the PDG-reported value: 1.06~cm$^{-2}$ min$^{-1}$ \cite{pdg}.

\begin{table}[!ht]
    \caption{Parameter values used in Eq. \ref{costheta}.
    \label{tab:my_label}}
    \centering
    \begin{tabular}{c|c}
    \hline\noalign{\smallskip}
         Parameter & Value\\
    \noalign{\smallskip}\hline\noalign{\smallskip}
         $P_1$ & 0.102573\\  
         %\hline
          $P_2$ & -0.068287\\ 
        %\hline
          $P_3$ & 0.958633 \\
        %\hline
          $P_4$ &  0.0407253\\
       % \hline
          $P_5$ & 0.817285\\
          %\hline
    \hline\noalign{\smallskip}      
    \end{tabular}
    \end{table}

As shown in Fig.~2 of Ref.~\cite{cosmugen}, this parametrization reproduces quite well the existing measurements \footnote{It is worth noticing that the most part of the world data are for $E_\mu >$10 GeV and only few are reported for lower energies.}.
X and Y coordinates of  cosmic muon tracks have been generated according to a uniform distribution on a plane crossing the geometrical center of the detector (Z=0), each track with an angular and energy dependence according to the model described above.  The generation plane was chosen  larger than the footprint of the detector (300 cm x 300 cm).  To take into account any possible material surrounding the telescope implemented in the GEANT model (concrete wall, roof, nearby buildings), once the track is generated, the vertex is moved far upwards,  ($\sim$80  m away), well outside any volumes present in the simulation \footnote{The shift applied to the vertex is proportional (x 0.8) to the size of the GEANT {\it World} volume that in this case was chosen as a cube of 100 m sides.}.
\begin{figure}[thb!]
\centering 
\includegraphics[width=0.9\columnwidth]{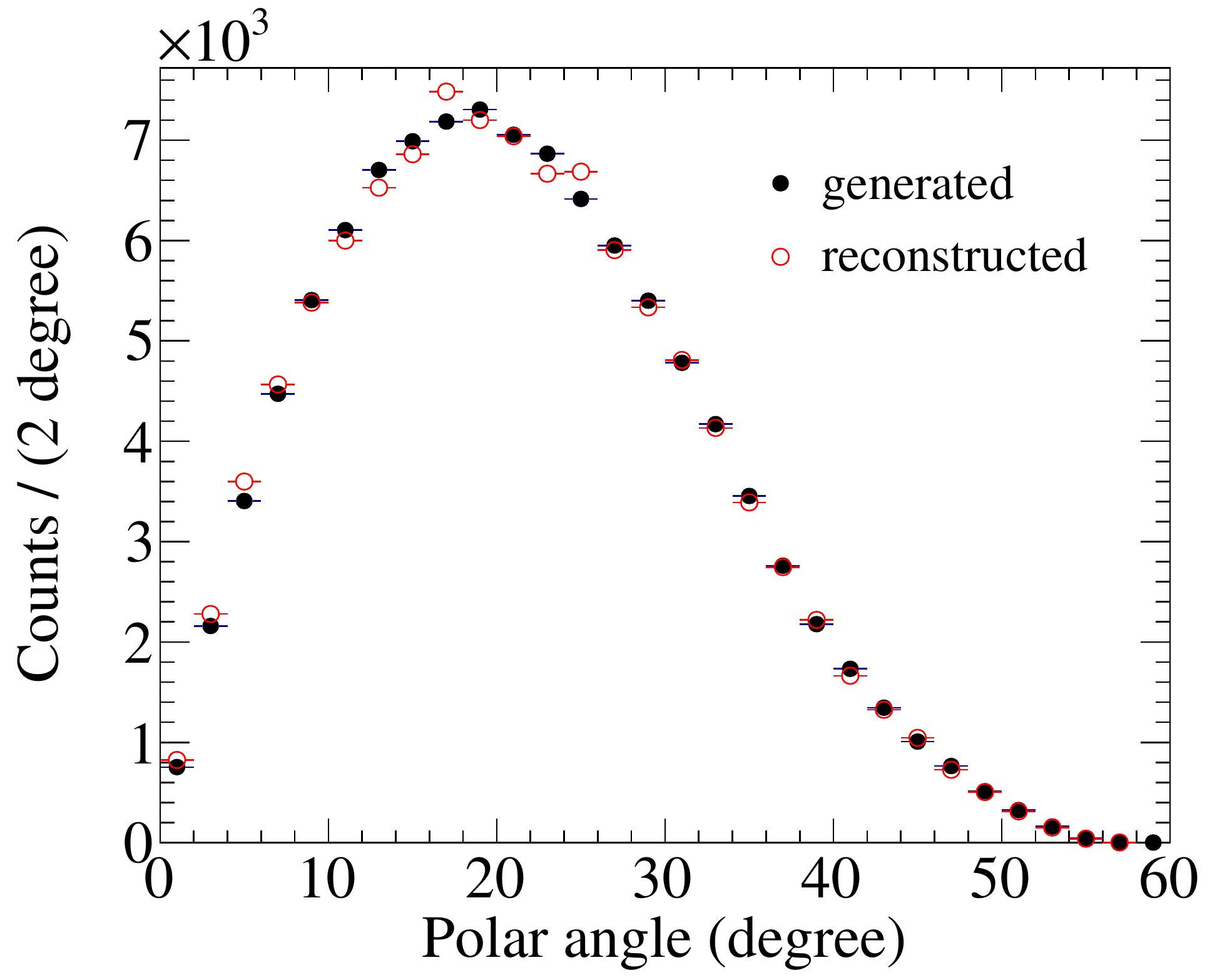}
\caption{Simulations results for the standard configuration telescope. Generated (full circles) and reconstructed (open circles) muon polar angle distribution. } 
\label{recon} 
\end{figure}
Generated muons were then passed to GEMC to obtain the EEE telescope expected angular distribution and counting rate of single-track events.
As an example, the generated and the reconstructed polar angle distributions are shown in Fig.~\ref{recon} for simulated cosmic muon in the energy range (0.2-100) GeV, for  a standard EEE telescope (50/50~cm). Detector acceptance and resolution effects are included. 

\section{Simulation framework validation}
\label{sim_frame_valid}
The EEE simulation framework has been validated by comparing MC predictions with data measured by EEE telescopes\footnote {All rates were corrected to take in account  altitude and atmospheric pressure variation. The correction procedure is the same described in Ref.~\cite{oulu}.}.

\subsubsection{Detector local efficiency}
\label{detect_eff}
Despite a detailed description of geometry and materials, the simulation framework described here uses a parametric response of the detector to impinging cosmic muons (see Sec.~\ref{detc_rep}). Details about the development of the ionization avalanche 
generated by the particle in the gas are missing. In real detectors 
the avalanche behaviour is a critical factor that strongly depends on parameters of the gas (composition, density, and flow), chamber geometry, operating voltage, and so on. They affect the signal characteristics, amplitude and time-over-threshold, determining the MRPC's spatial and temporal resolution, and determine the detection efficiency.
It is therefore necessary, before comparing data to simulations, to extract a precise map of the MRPCs efficiency and correct raw data. In the following we show how the efficiency correction is estimated and applied to data.

The MRPCs efficiency can be evaluated using some reference detectors, e.g. movable scintillator counters located on top and bottom of each MRPC in time-coincidence with the impinging cosmic muon. However, the complications related to in-situ measurement of each telescope, the time necessary for a complete scan of the three MRPCs and the difficulty in repeating the efficiency assessment over time, suggested a different approach.
Making use of the track measurement redundancy provided by the three chambers (two can be used as a reference to determine the efficiency of the third) we developed a method to 
map out the efficiency of individual MRPCs of the EEE telescope. The procedure uses the same data sample selected for physics analyses, providing a map that correctly takes into account any local efficiency variation in any time period under analysis.

\paragraph{The procedure.}
The MRPCs efficiency has been computed by subdividing the active area 
of a chamber in 20$\times$24 bins, 7.9~cm $\times$ 3.2~cm each, 
corresponding to X$\times$Y coordinates, respectively. In each bin we combined a {\it tracking} efficiency, that takes into account the 
probability of reconstructing a good track (all detected hits close to the track candidate), with a {\it counting} efficiency that takes into account the non-uniform hit distributions on the chamber's 
surface due to non-homogeneous MRPC response.
\begin{figure}[h!]
\centering 
\centering
\includegraphics[width=0.9\columnwidth]{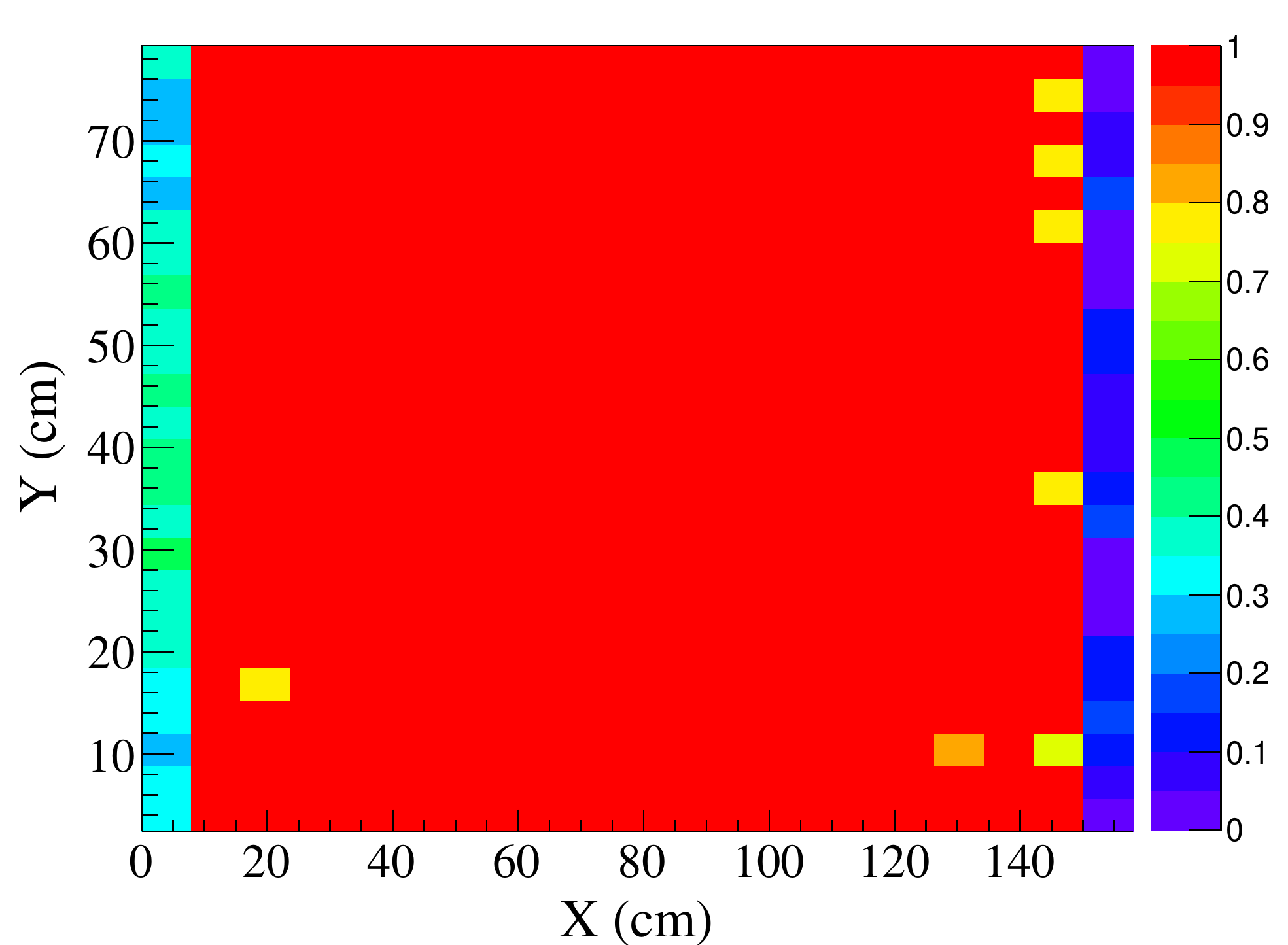}
\includegraphics[width=0.9\columnwidth]{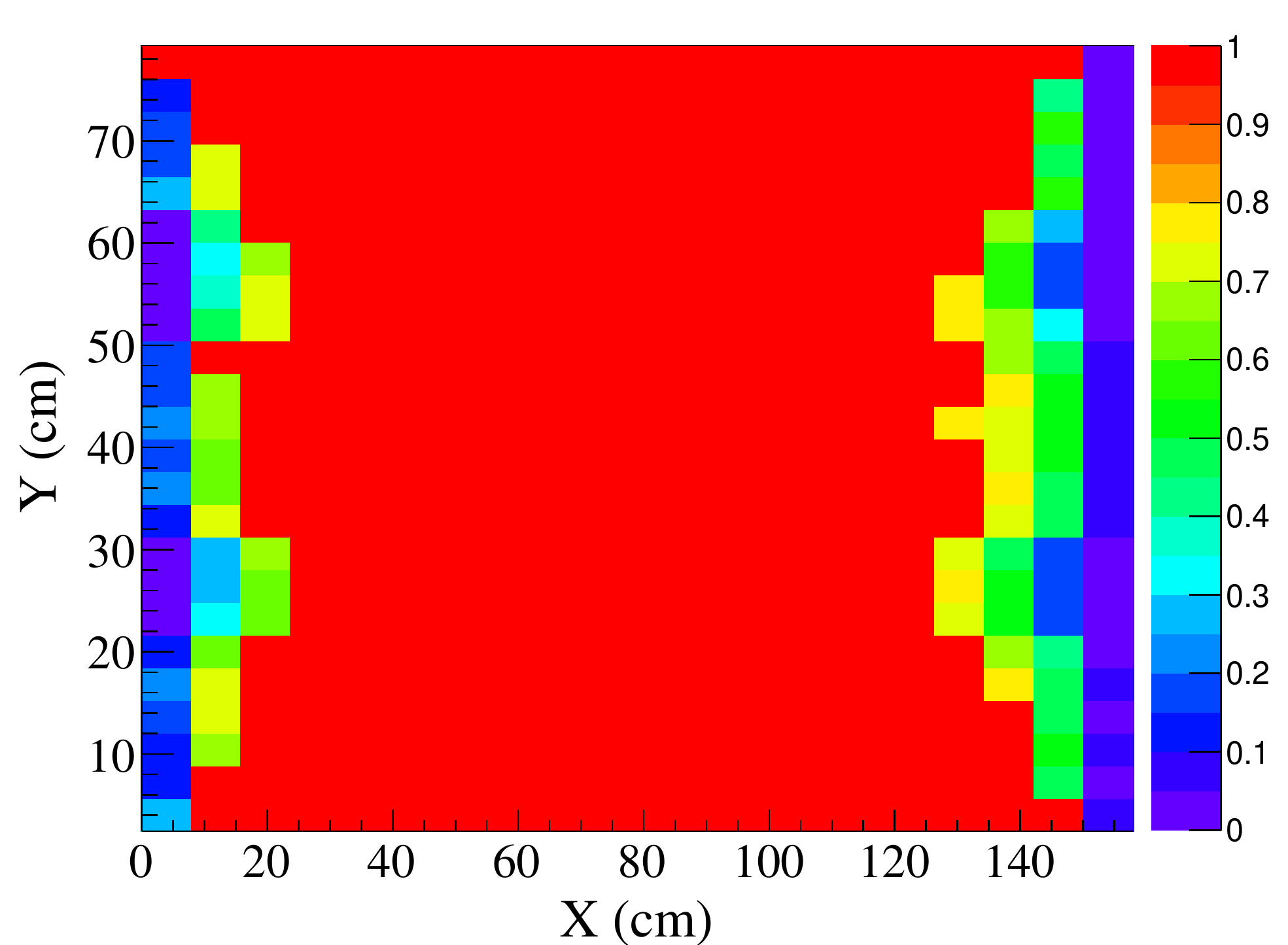}
\includegraphics[width=0.9\columnwidth]{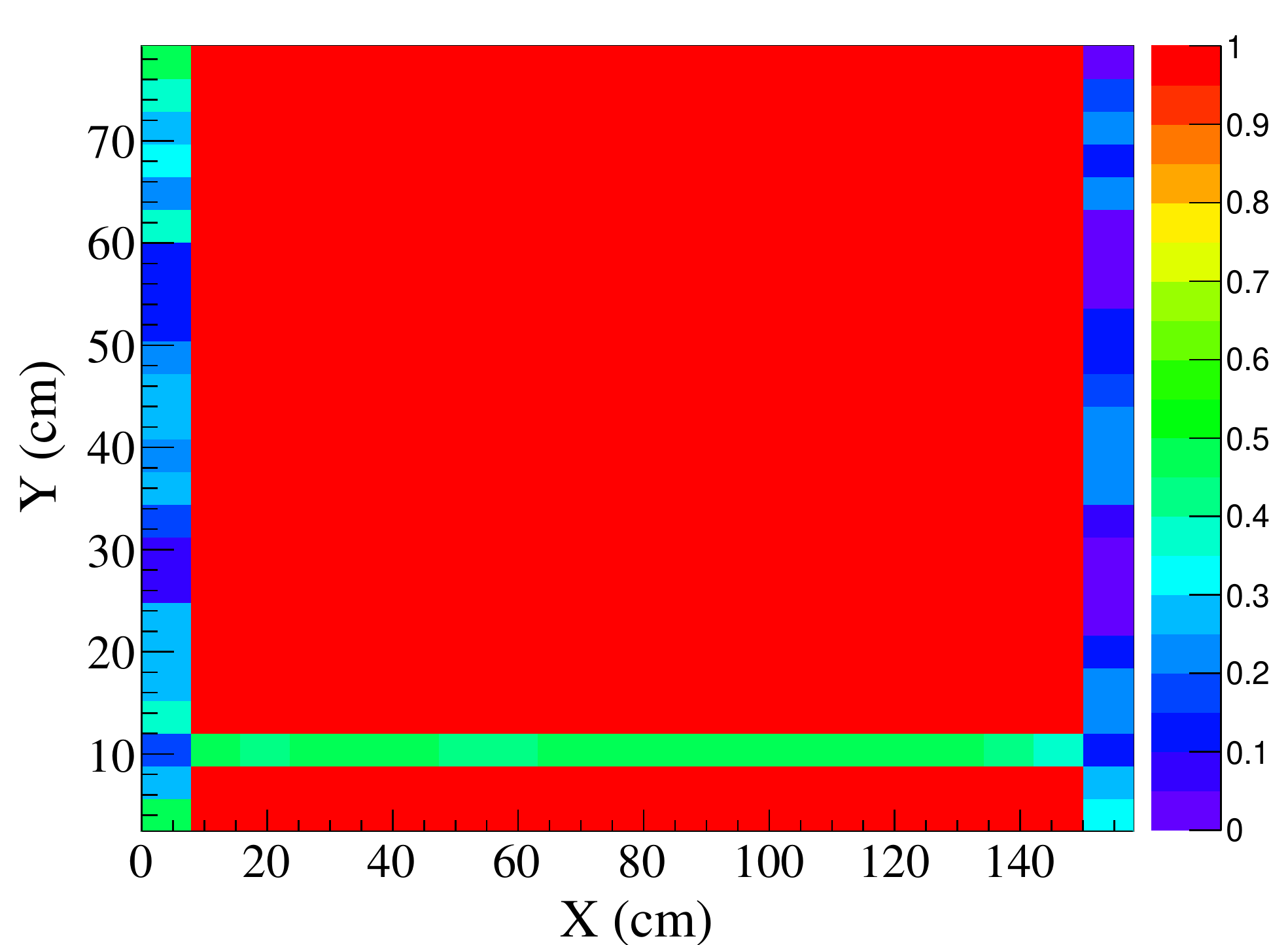}
\caption{Local efficiency maps for the three MRPCs composing the TORI-03 telescope. The three maps correspond to the top, middle, and bottom chambers, respectively.\label{effitori3mandy} } 
\end{figure}
To determine the {\it tracking} efficiency 
regular events triggered by hits in all three chambers are considered. The chamber under examination is ignored: only hits from the other two chambers are used in the fit, deriving a track that is then projected onto the corresponding (X,Y) bin of the third chamber. If a hit is present inside that bin, the hit is classified as ``detected'' and a corresponding map is filled. If it is not, the hit is classified as ``missing'' and the hit is stored in a different map. In this last case the recorded hit is triggered by electronic noise in the chamber since the trigger condition
requires a three-fold coincidence across all the chambers in the telescope. The {\it tracking} efficiency map is obtained as the ratio between the maps of {\it detected} and {\it detected + missing} hits. 
Since the procedure does not depend on the track angle, for simplicity, only vertical tracks were used.

For each chamber, the counting efficiency  (X,Y) map is obtained as the ratio of data and simulation hit distributions  normalized to the average value of the most uniform area. In this way the relative inefficiency, with respect to areas  where the detector is performing as expected, is taken into account.
The absolute efficiency (local efficiency) map is obtained as the product of the {\it counting} and  {\it tracking}  efficiency maps. More details about the procedure are reported in Ref.~\cite{mandy}.
As a cross-check, for selected bins, the resulting efficiency was compared, to the efficiency obtained by using two external plastic scintillator counters. The two methods provided consistent results within the experimental errors.

The local efficiency maps obtained in this way for the three chambers of TORI-03 are shown in Fig.~\ref{effitori3mandy}. 
This method is sensitive to single (X,Y) bin inefficiency (such as the ones visible, for instance, in the map of the top chamber), or inefficiency related to a reduced gas flow (most likely localized on the chamber sides as visible in the map of the middle chamber) or inefficiency corresponding to a missing full strip (visible in the map of the bottom chamber). To provide a good accuracy in the local efficiency map, at least the statistics corresponding to about 24h of data taking is needed.
\begin{figure}[th!]
\includegraphics[width=1.0\columnwidth]{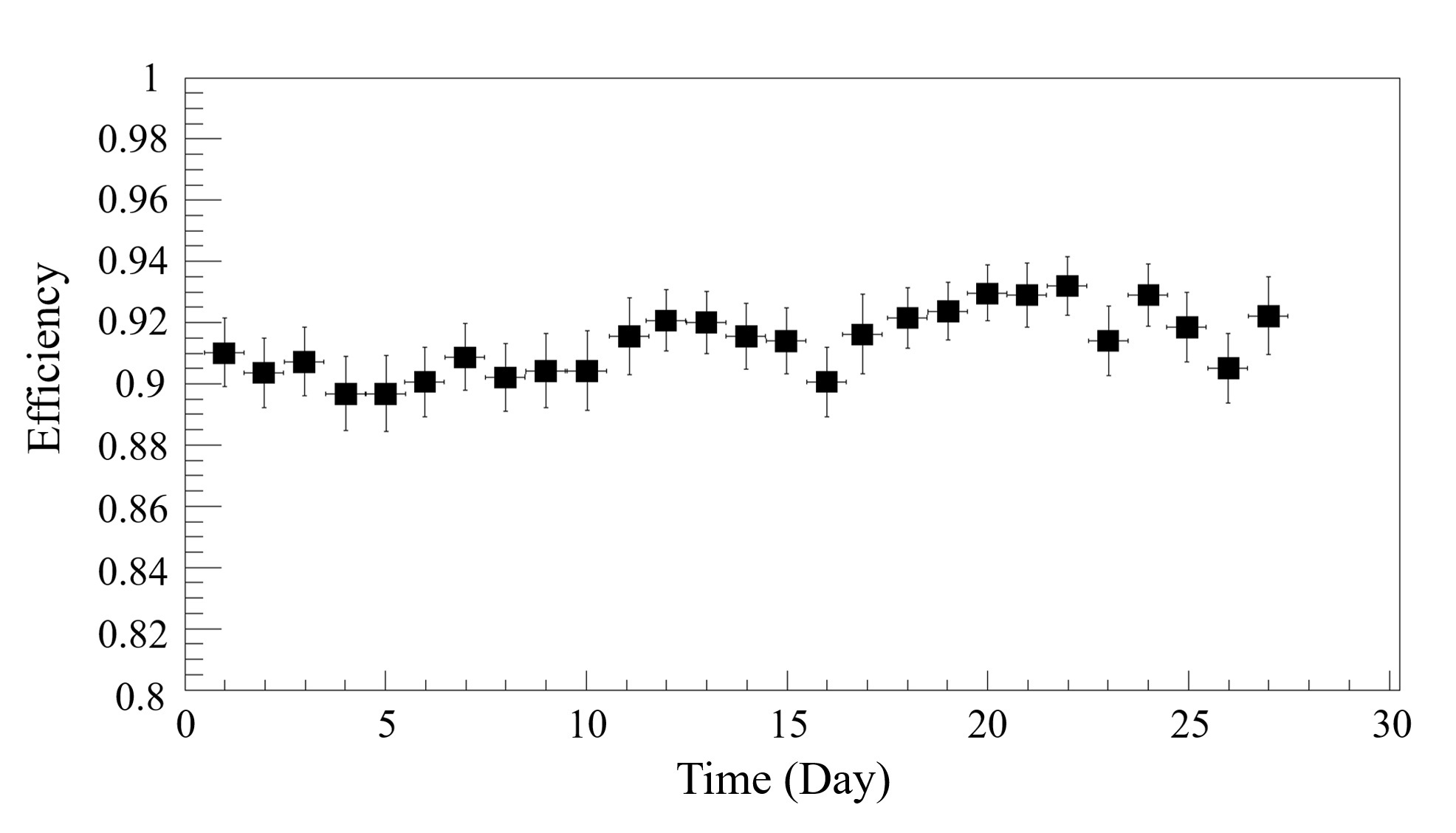}
\caption{Local efficiency integrated over X and Y for the TORI-03 middle chamber. Each point corresponds to one-day of data taking.} 
\label{fig:tori03-eff-t}
\end{figure}

The very same method can be used to monitor the variation over time of the local efficiency by comparing a map measured at a certain time during the data taking to a reference measured at the beginning.
The TORI-03 middle chamber efficiency as a function of time, obtained by integrating daily (X,Y) maps over the whole chamber area is shown in Fig.~\ref{fig:tori03-eff-t}.
\begin{figure}[th!]
\centering 
\includegraphics[width=.9\columnwidth]{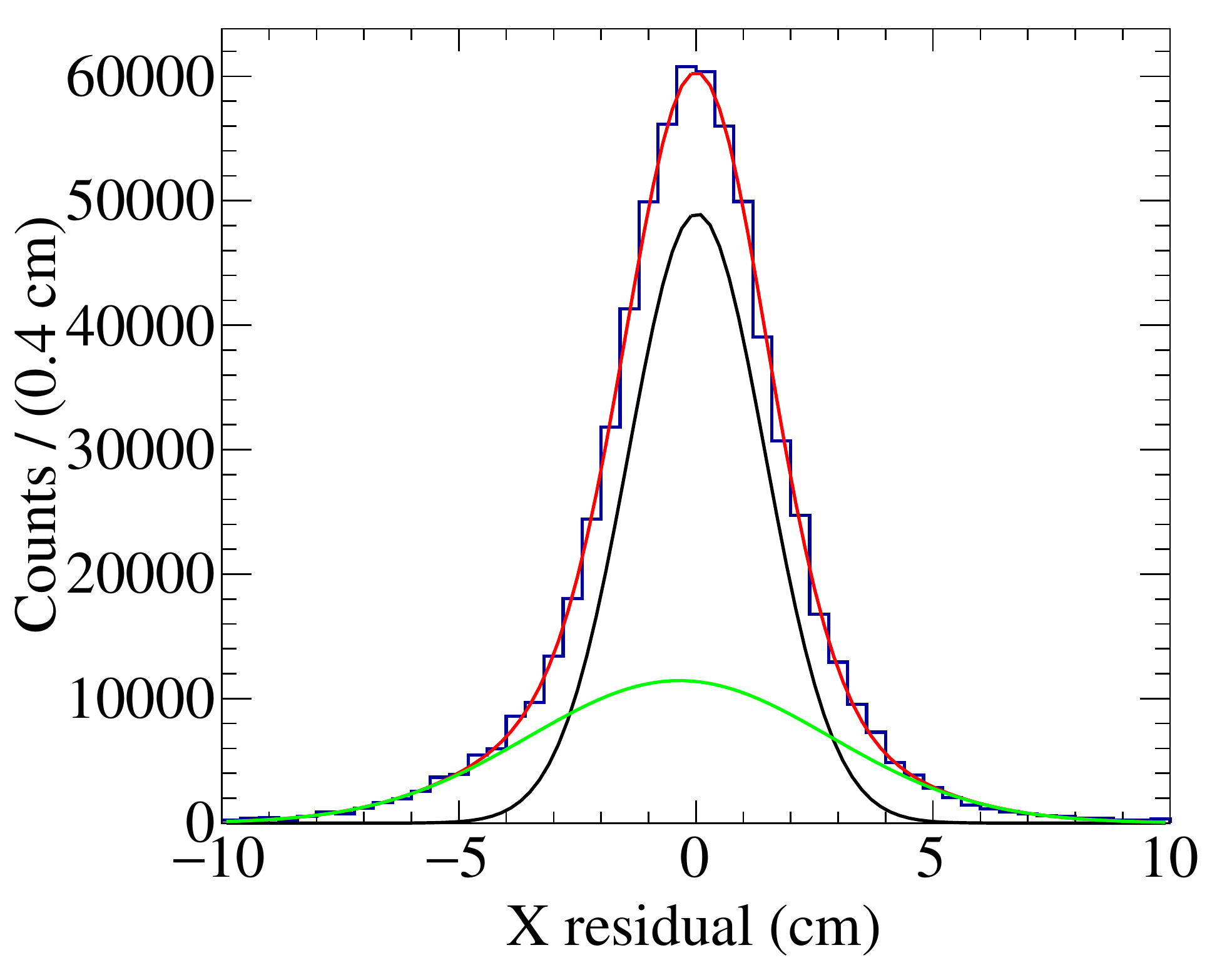}\\
\includegraphics[width=.9\columnwidth]{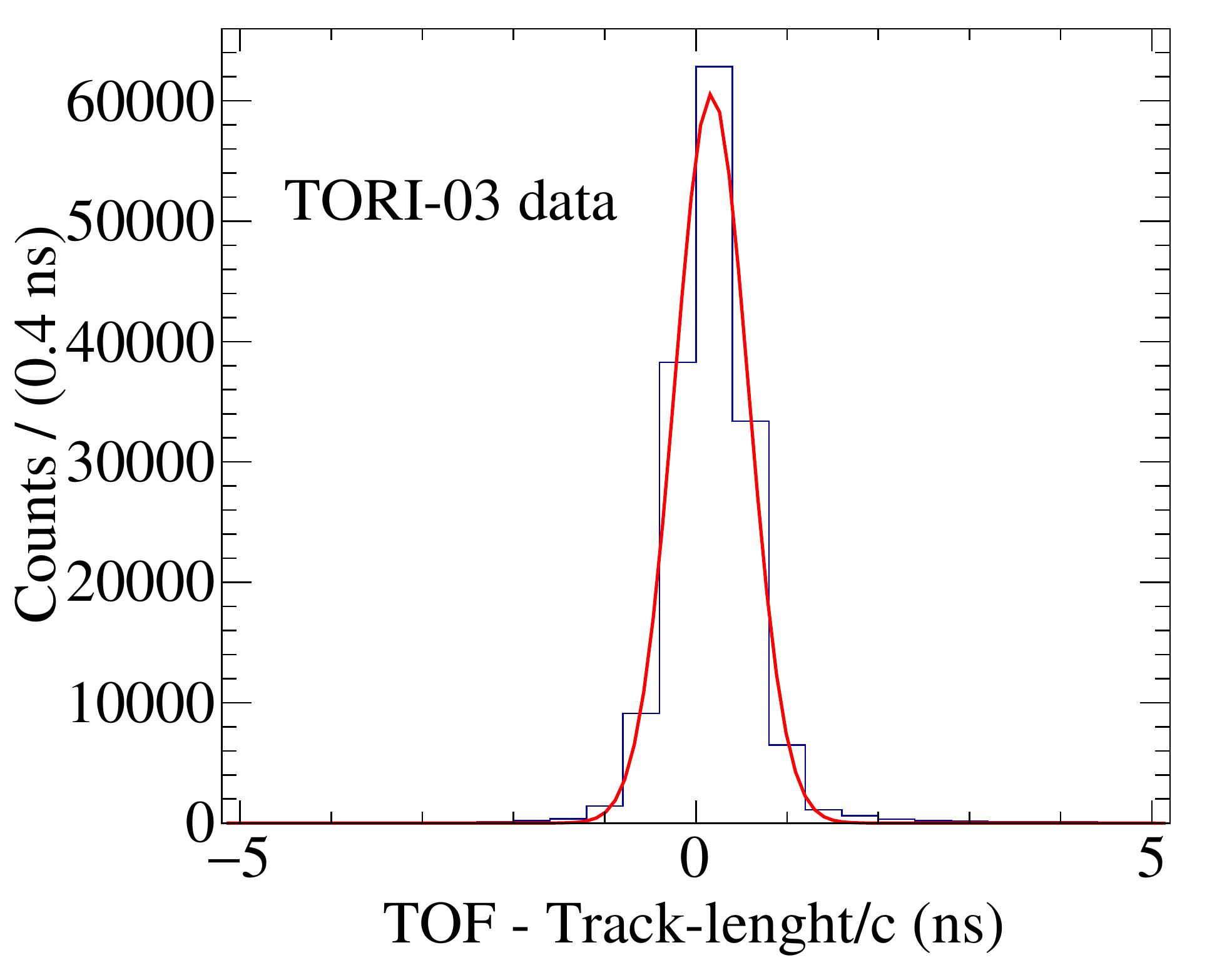}\\
\includegraphics[width=.9\columnwidth]{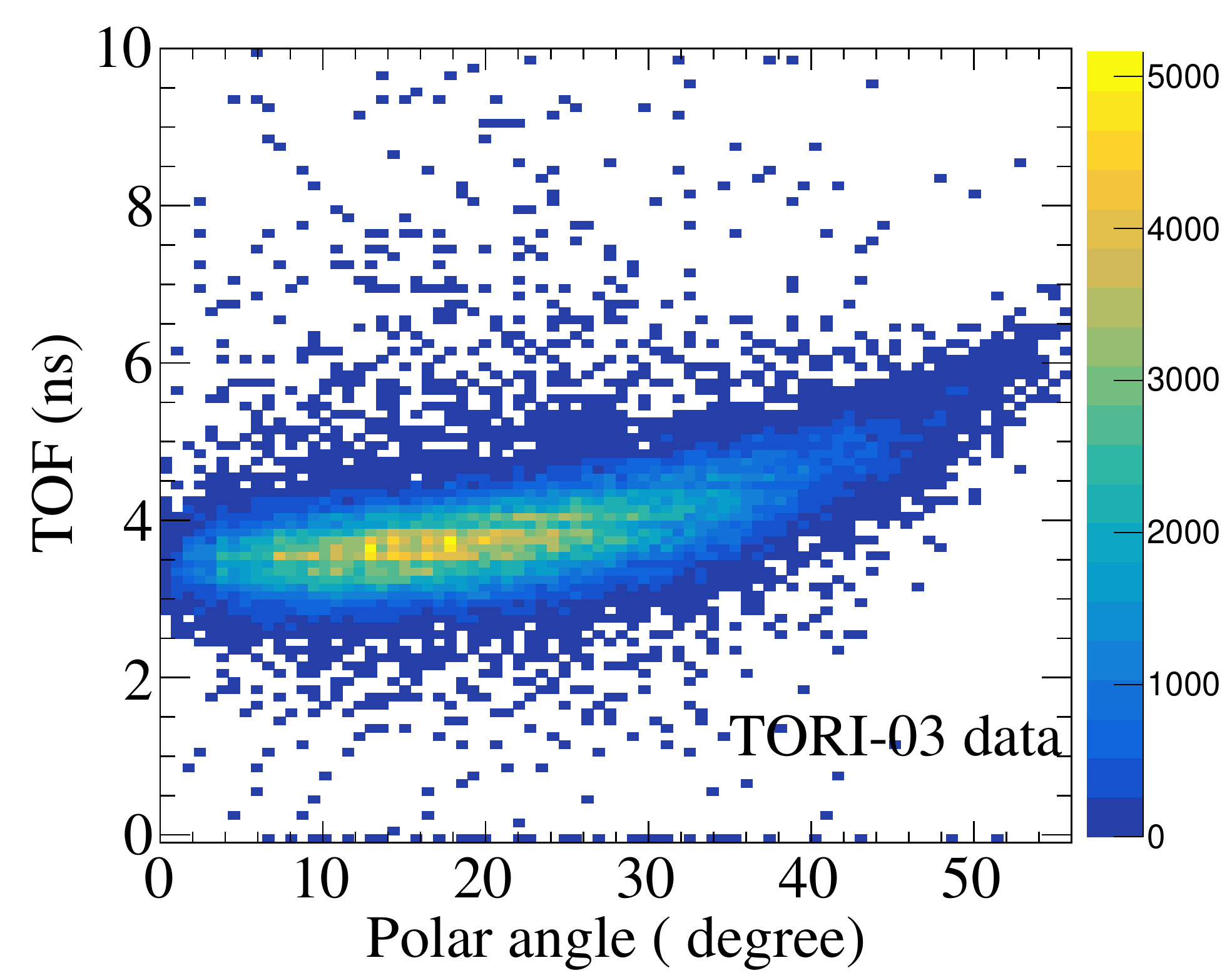}
\caption{TORI-03 telescope track X coordinate residual (top), Time-of-Flight - track length over $c$ (with respect to the speed of light) (middle), correlation between tracks Time-of-Flight and polar angle (bottom). In the top (middle) panel, the distribution is fit with a double (single) Gaussian, shown as a red curve. The standard deviation of the narrow Gaussian in the top panel is $\sigma_{\rm X}=(1.4\pm0.3$)~cm. The fit mean value and the standard deviation in the time distribution (middle panel) are ($0.16\pm 0.04$)~ns and ($0.37\pm0.03$)~ns, respectively.} 
\label{fig:Xinter-XPosMidTori}
\end{figure}

\subsection{Comparison between data and simulations}

\label{data_sim_comp}
The TORI-03 experimental set-up was % implemented in GEMC considering:
implemented in GEMC using the standard geometry and surrounding.
%\begin{itemize}
%\item a distance between chambers of 50/50~cm;
%\item a telescope within a concrete box with 30-cm-thick side walls and roof; 
%\item no surrounding buildings or obstacles.
%\end{itemize}
Experimental data have been corrected  by weighting each event with the corresponding value of the local efficiency map.

\begin{figure}[th!]
\centering 
\includegraphics[width=.9\columnwidth]{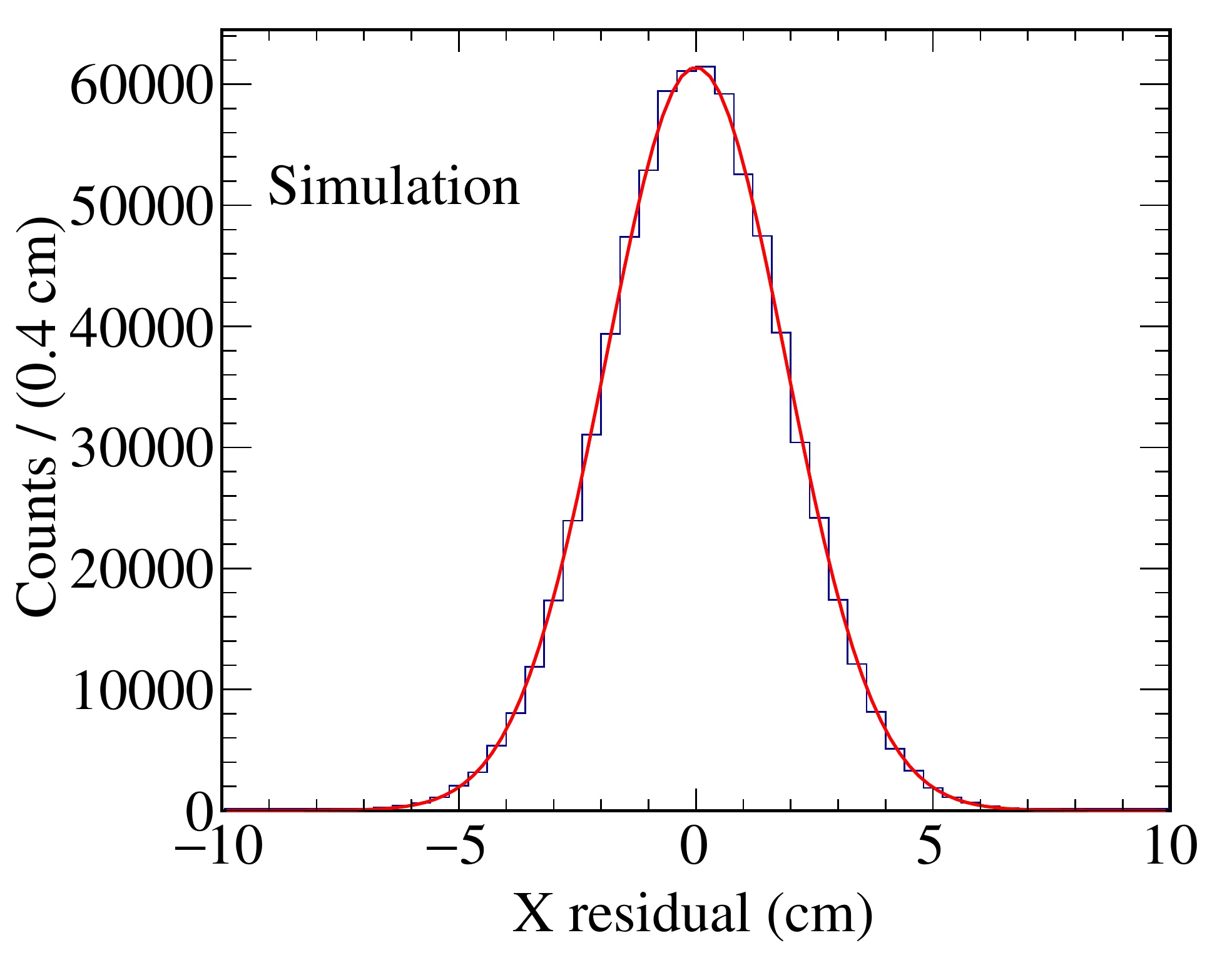}
\includegraphics[width=.9\columnwidth]{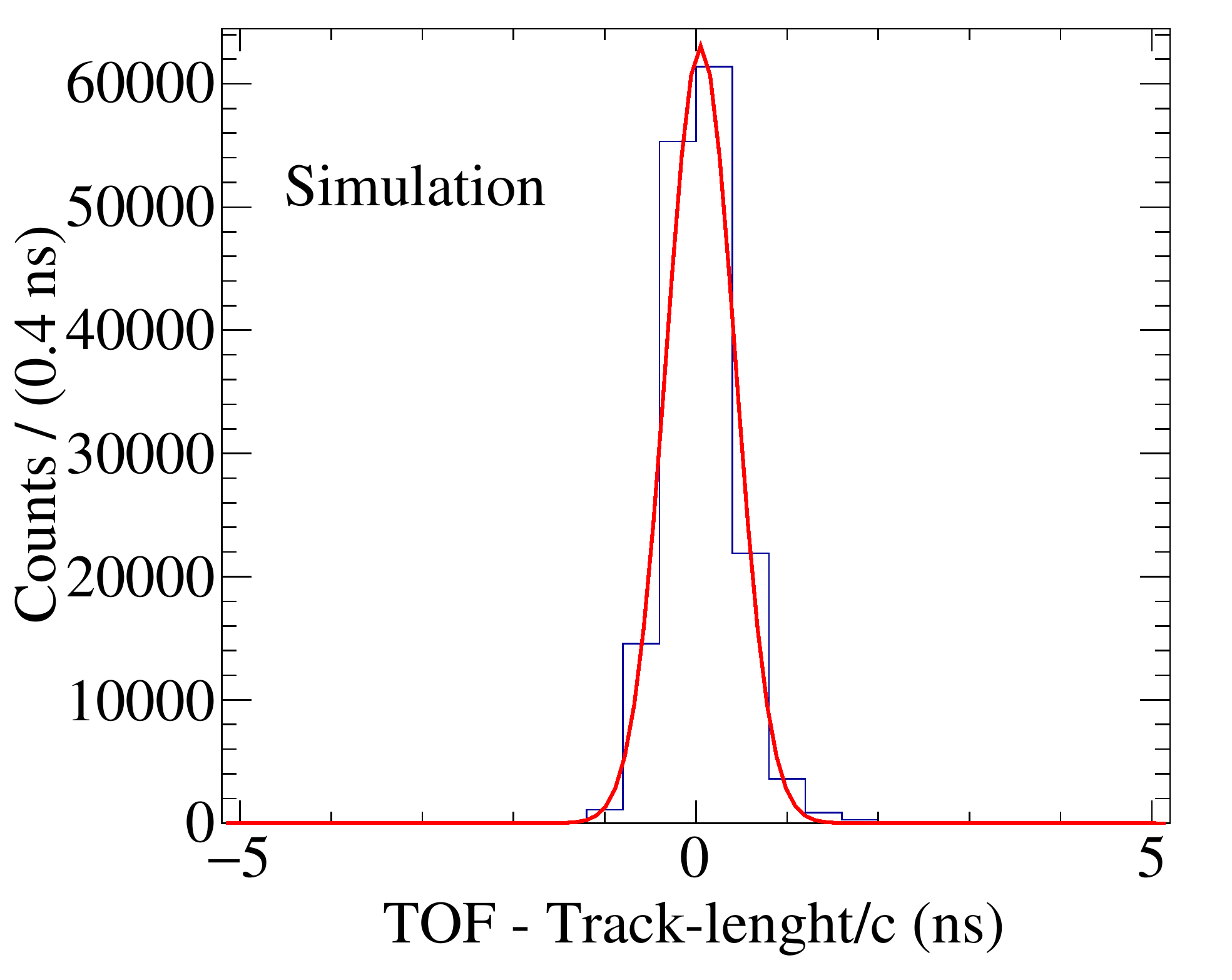}
\includegraphics[width=.9\columnwidth]{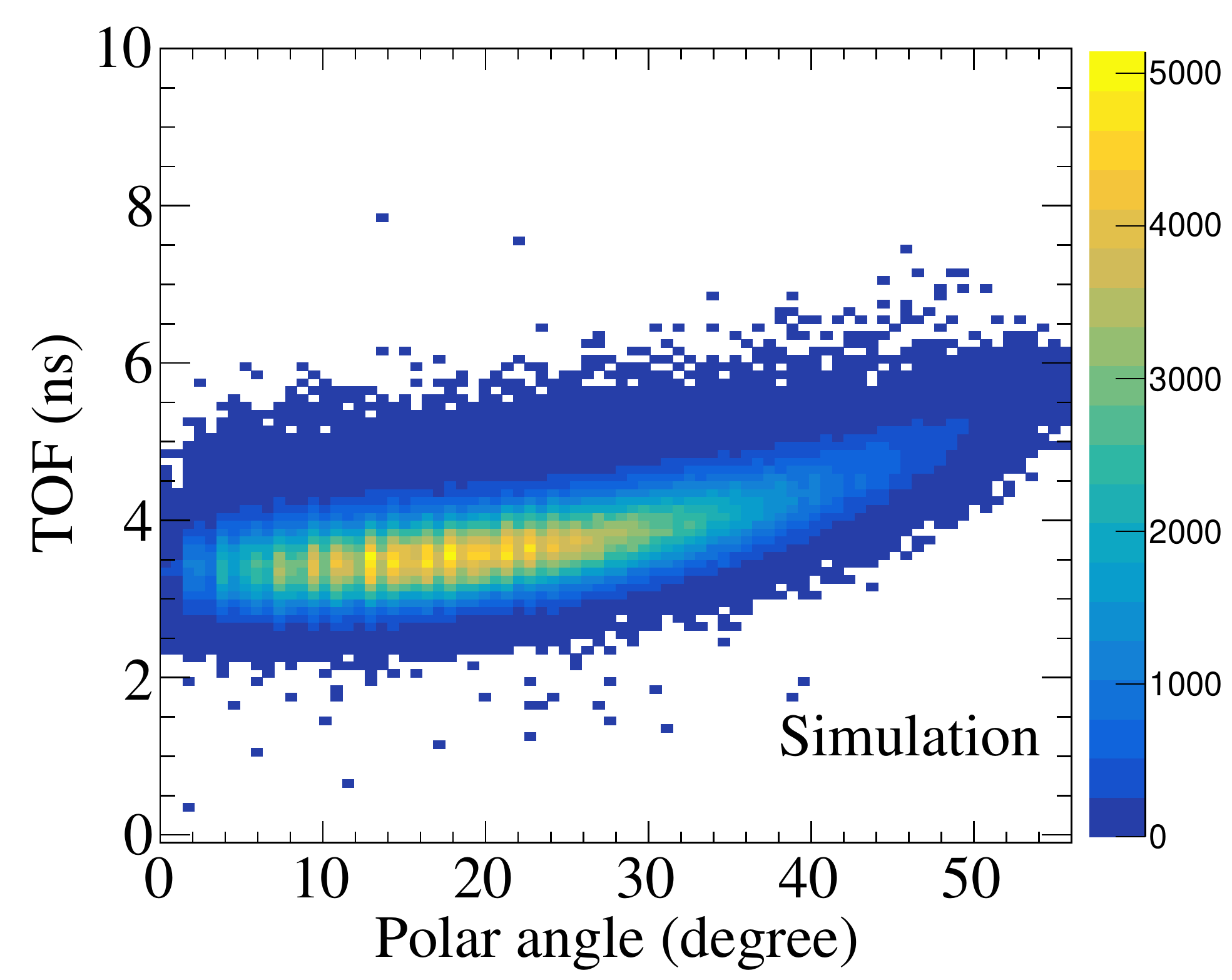}
\caption{TORI-03 telescope simulation. Observables are the same as in figure~\ref{fig:Xinter-XPosMidTori}. The standard deviation of the Gaussian fits are: $\sigma_{\rm X}=(1.90\pm0.15$)~cm, $\sigma_{\rm TOF-TrL}=(0.40\pm0.05$)~ns, respectively; the mean value of the Gaussian fit of the time distribution (middle panel) results ($0.10\pm0.03$) ns.} 
\label{fig:Xinter-XPosMidSIM}
\end{figure}

Both the parametrization of detector response and the model used to generate cosmic muons contribute to the reported observables.
Figures~\ref{fig:Xinter-XPosMidTori} and ~\ref{fig:Xinter-XPosMidSIM} show the comparison of some observable derived from experimental data (Fig.~\ref{fig:Xinter-XPosMidTori}) and simulations (Fig.~\ref{fig:Xinter-XPosMidSIM}). The top panel shows the X coordinate residual (with respect to the fitted track) for the telescope middle chamber.
The middle graph shows the Time-of-Flight residual
calculated as the difference between the measured value (deduced by the time difference between top and bottom hit time) and the time obtained by the measured track length divided by $c$. 
These quantities are assumed to be estimators of spatial and time resolution, respectively.
The little shift from zero of the Time-of-Flight residual (($0.16\pm0.04$)~ns for data and ($0.10\pm0.03$)~ns for simulation) is due to low momentum muons with speed lower than $c$. 
The bottom panel shows the correlation between the track Time-of-Flight and the polar angle. Distributions from data and simulations are quite similar from a qualitative and quantitative point of view.
\begin{figure}[th!]
\centering 
\includegraphics[width=.95\columnwidth]{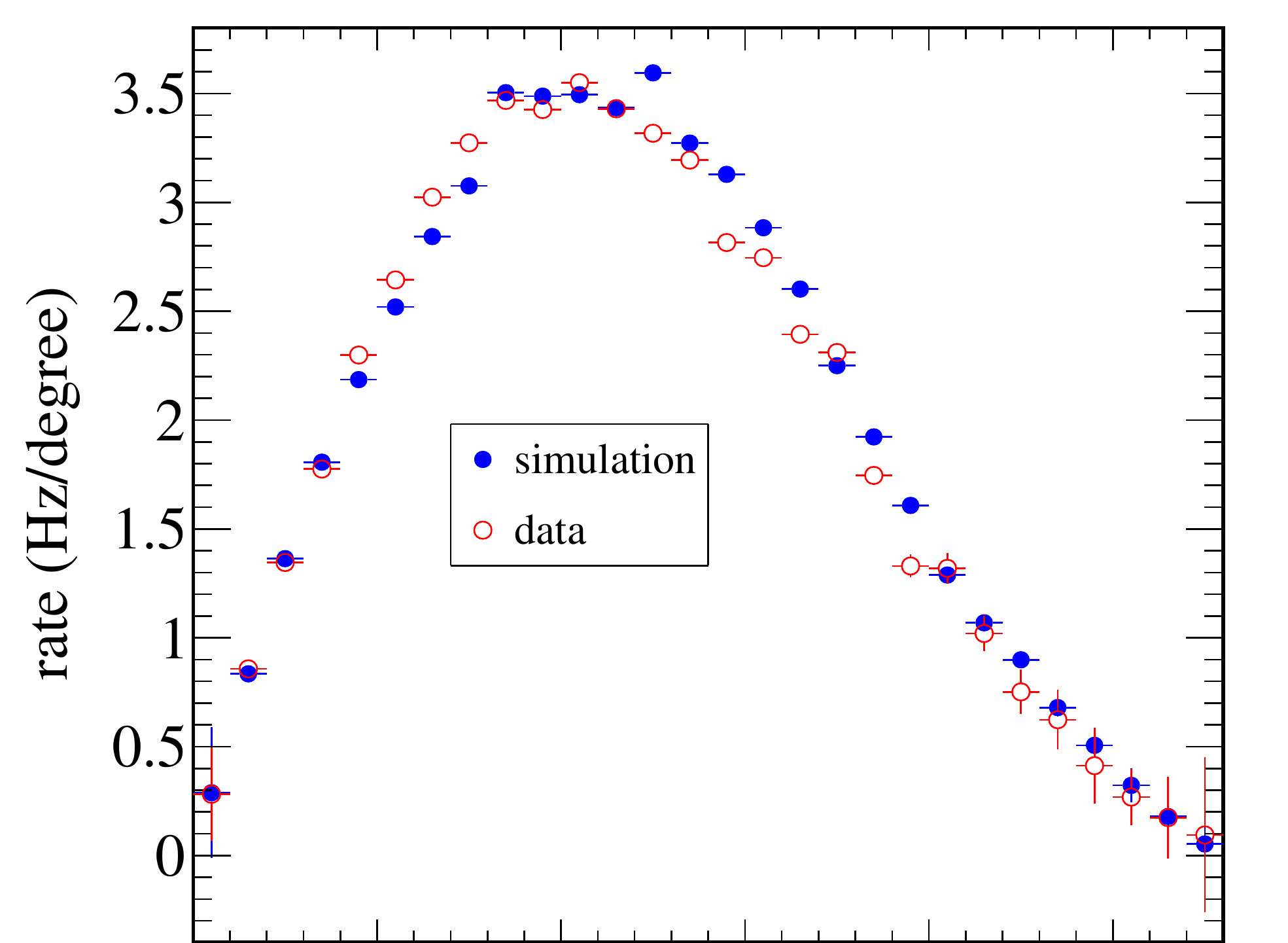} 
\includegraphics[width=.95\columnwidth]{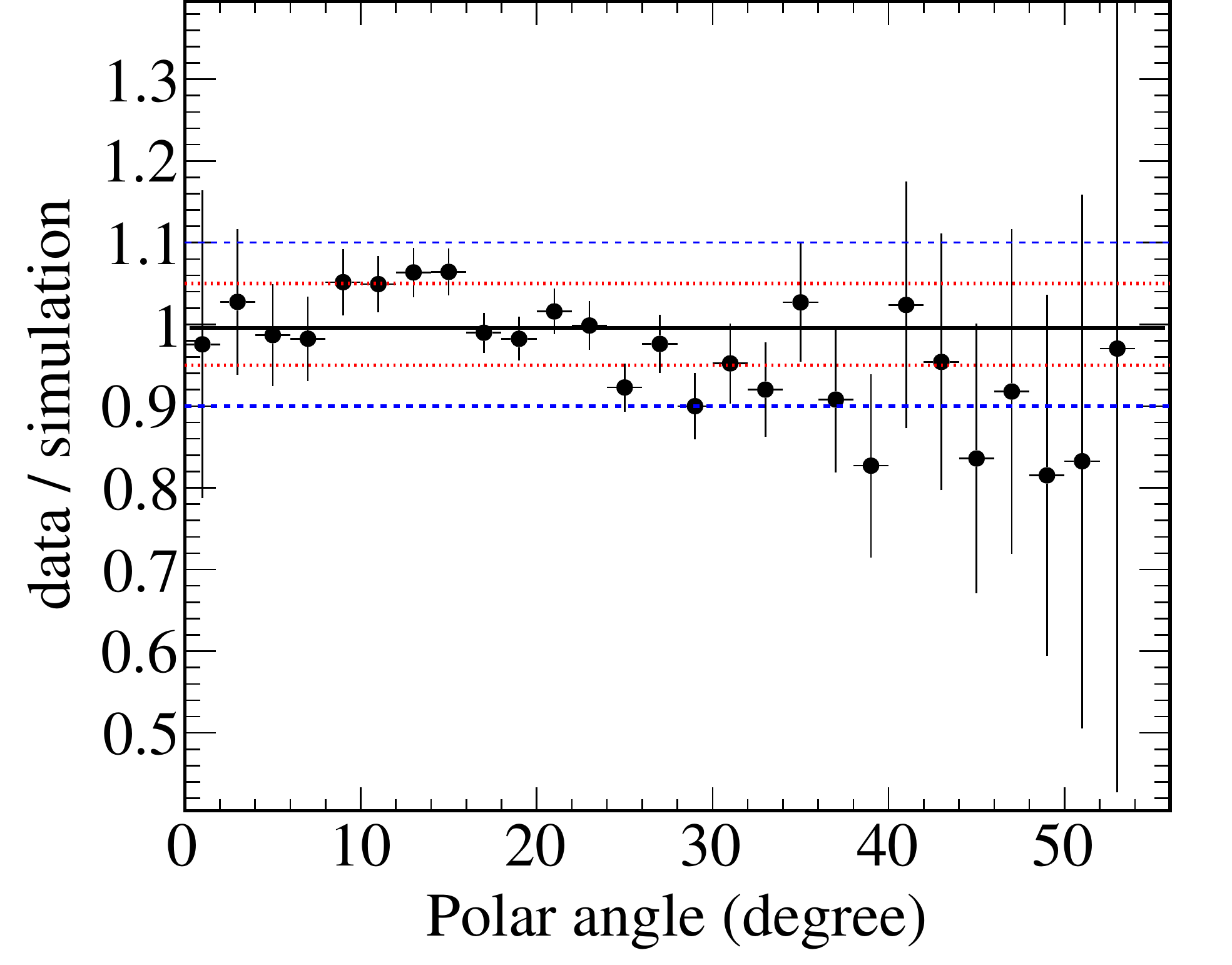}
\caption{Comparison between cosmic muon rate as a function of the track polar angle for TORI-03 data (red open circles) and the ones predicted by simulations (blue full circles). The ratio between the two distributions is shown in the bottom panel. The red (blue) line delimits a $\pm5\%$ ($\pm10\%$) band, the solid (black) line corresponds to a fit with a constant function found to be 1.00$\pm$0.01.
\label{fig:torirate}}
\end{figure}

The comparison between single-muon rates measured by TORI-03 and the ones predicted by simulations, as a function of the muon track polar angle, is shown in Fig~\ref{fig:torirate}. 
The top panel shows the two distributions superimposed, while the bottom panel shows their ratio. 
The good agreement at all angles and the mean value of the ratio, resulting to be around unity, demonstrate that the EEE simulation framework is able to reproduce the absolute observed angular cosmic muon rate within an error of about 5$\%$ for polar angle lower than 25$^\circ$ and of about 10\% for polar angle lower than 38$^\circ$.
Similar agreement was obtained for the CERN-01 telescope.% \textcolor{blue}{, one of the first built telescope of the EEE network, well maintained and with stable performance over time} \textbf{Al solito, ho aggiunto questa specifica rispetto a quanto suggerito dal referee, si potrebbe omettere, che suggerite?}.
%The EEE CERN-01 telescope is one of the first built telescope of the EEE network, well maintained and with stable performance over time.Its local efficiency map  shows a high and uniform efficiency, close to 100$\%$, for the most part of the chambers active area. With a similar procedure  we calculated the ratio between the distributions of track polar angle measured by CERN-01 and simulation. The result, compatible with 1  for all angles, confirms again the reliability of the EEE simulation framework. 

As a final validation of the EEE simulation framework, Tab.~\ref{tab:rates} reports the comparison between the integrated cosmic muon rate measured by TORI-03 and CERN-01 telescopes and the ones predicted by simulations. 
The good agreement, within few percent, demonstrates that the EEE simulation framework reproduces well the experimental data and can be used to compare telescopes featuring various experimental setups.
\begin{table}
\begin{center}
{\small
\begin{tabular}{|c|c|c|} \hline
	EEE telescope & Data Rate (Hz) & Simulated Rate (Hz)\\
\hline\hline
   TORI-03 & $ 54 \pm 5 $\ & $ 55 \pm 5$\\\hline 
   CERN-01 & $ 57 \pm 5 $\ & $ 58 \pm 5$\\ \hline 
\end{tabular}}
\caption{Cosmic muon rates measured by two EEE telescopes and corresponding values predicted by simulations. Data rates are corrected to take into account the local efficiency (as described in the text), simulation rates are corrected to take into account the elevation of the two telescopes. The data rate error include a systematic error of $8\%$ that accounts for the fluctuations of the operating conditions of the EEE telescopes in time. }
\label{tab:rates}
\end{center}
\end{table}

\section{Systematic studies}
\label{eee_tel_sim}
In this section we report some results obtained by means of the simulation, which provide a deeper understanding of the telescope performance, in particular for what concerns angular and spatial resolution, detection efficiency, and the effects of materials surrounding the telescope. %on its performance.
\begin{figure}[h!]
\centering 
\includegraphics[width=.9\columnwidth]{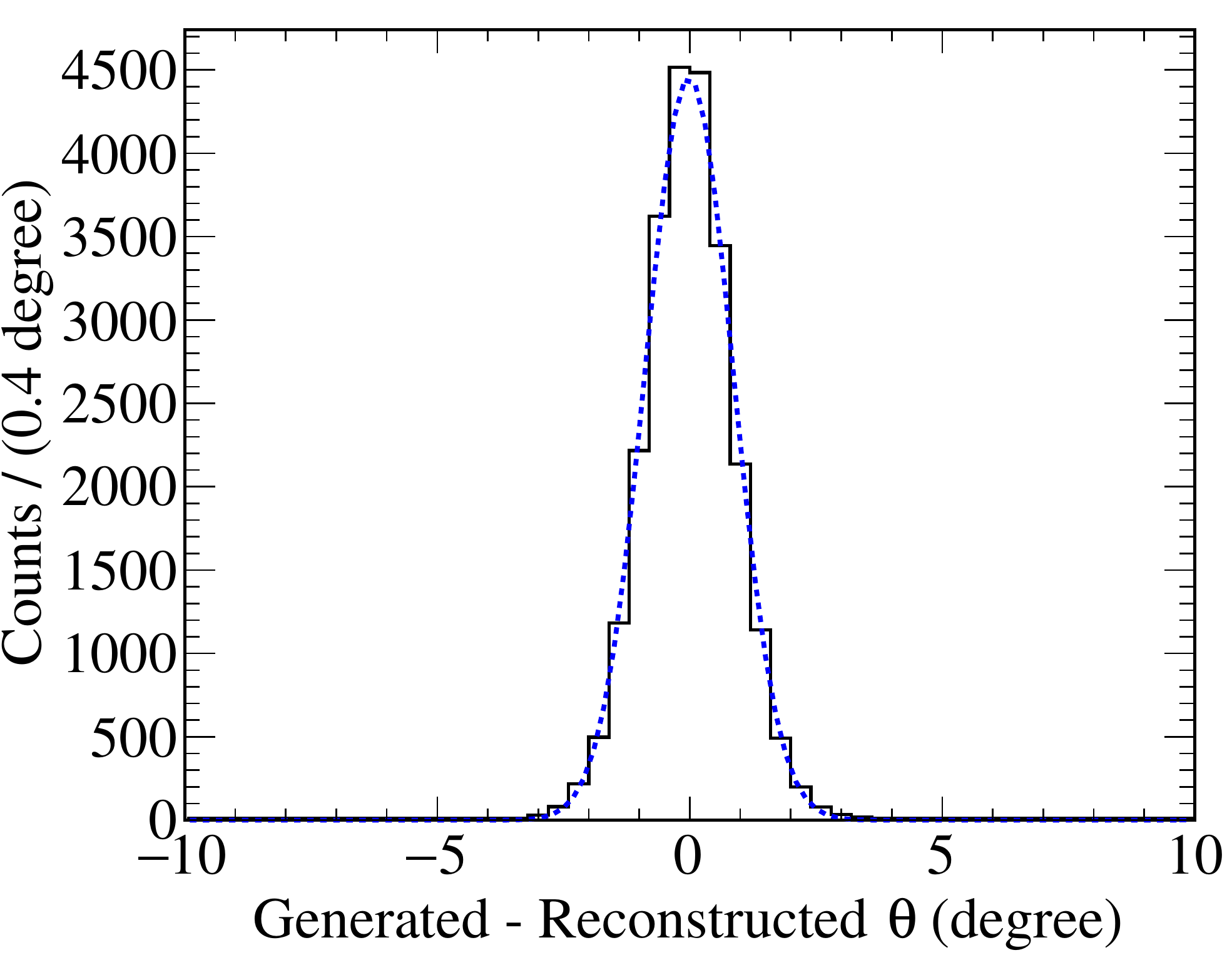} 
\caption{EEE telescope angular resolution defined as difference between generated and reconstructed polar angle for 10 -100 GeV energy muons. Full line is simulated data, dashed line the result of a Gaussian fit.
\label{reshig1} } 
\end{figure}

\subsection{Resolution}
\label{detect_res_condition}
\begin{figure}[h!]
\centering 
\centering
\includegraphics[width=0.9\columnwidth]{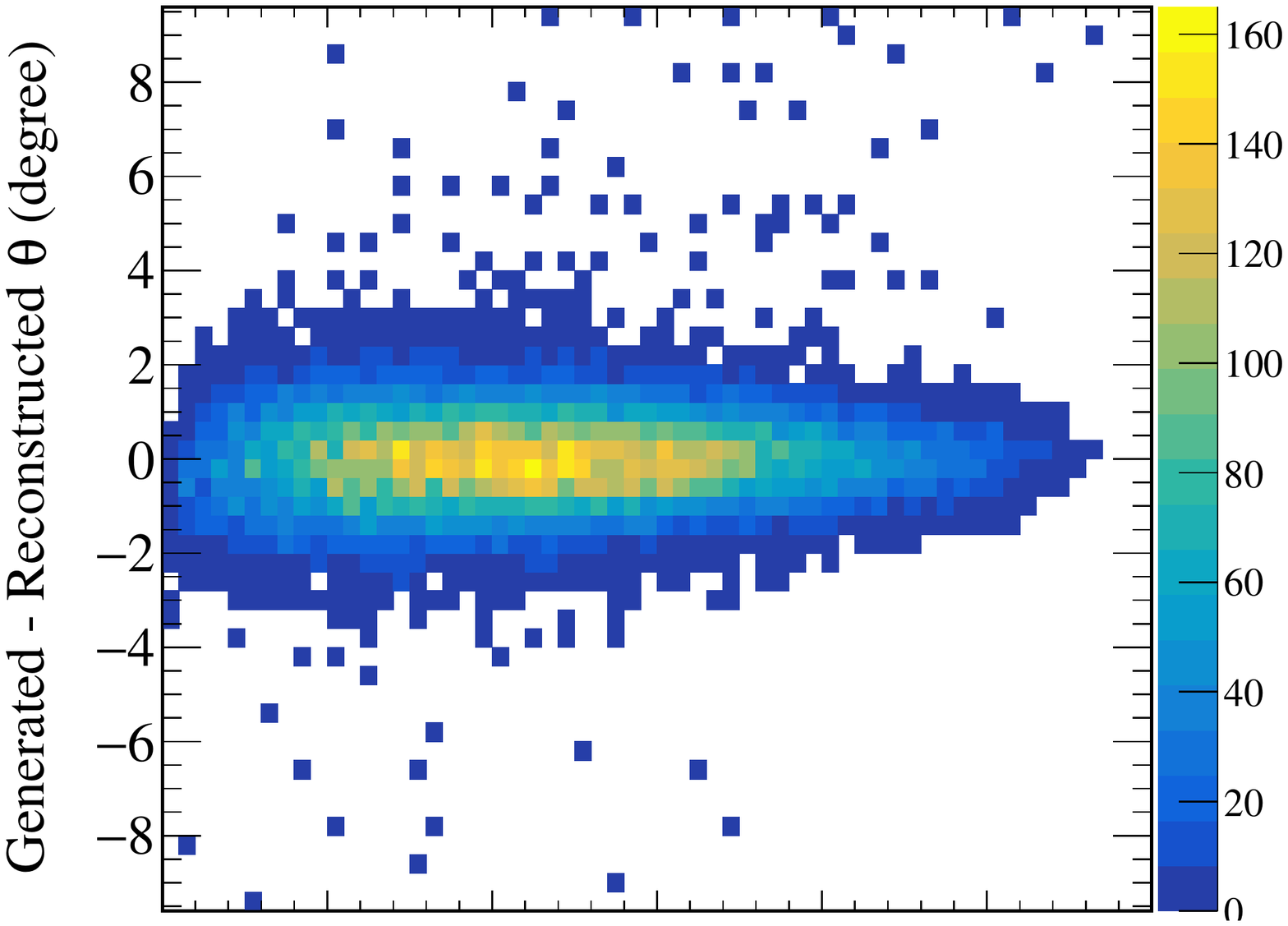}
\includegraphics[width=0.9\columnwidth]{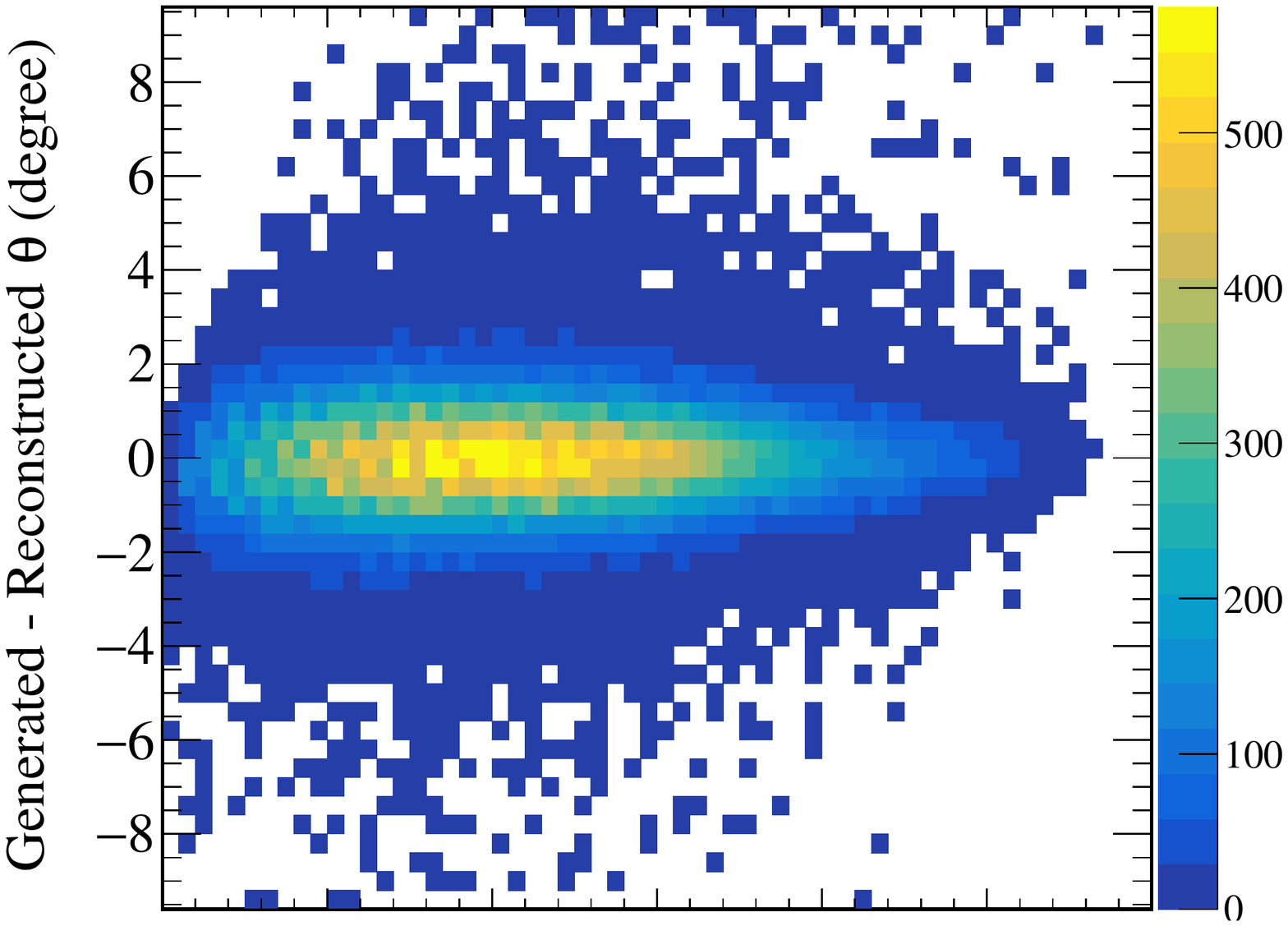}
\includegraphics[width=0.9\columnwidth]{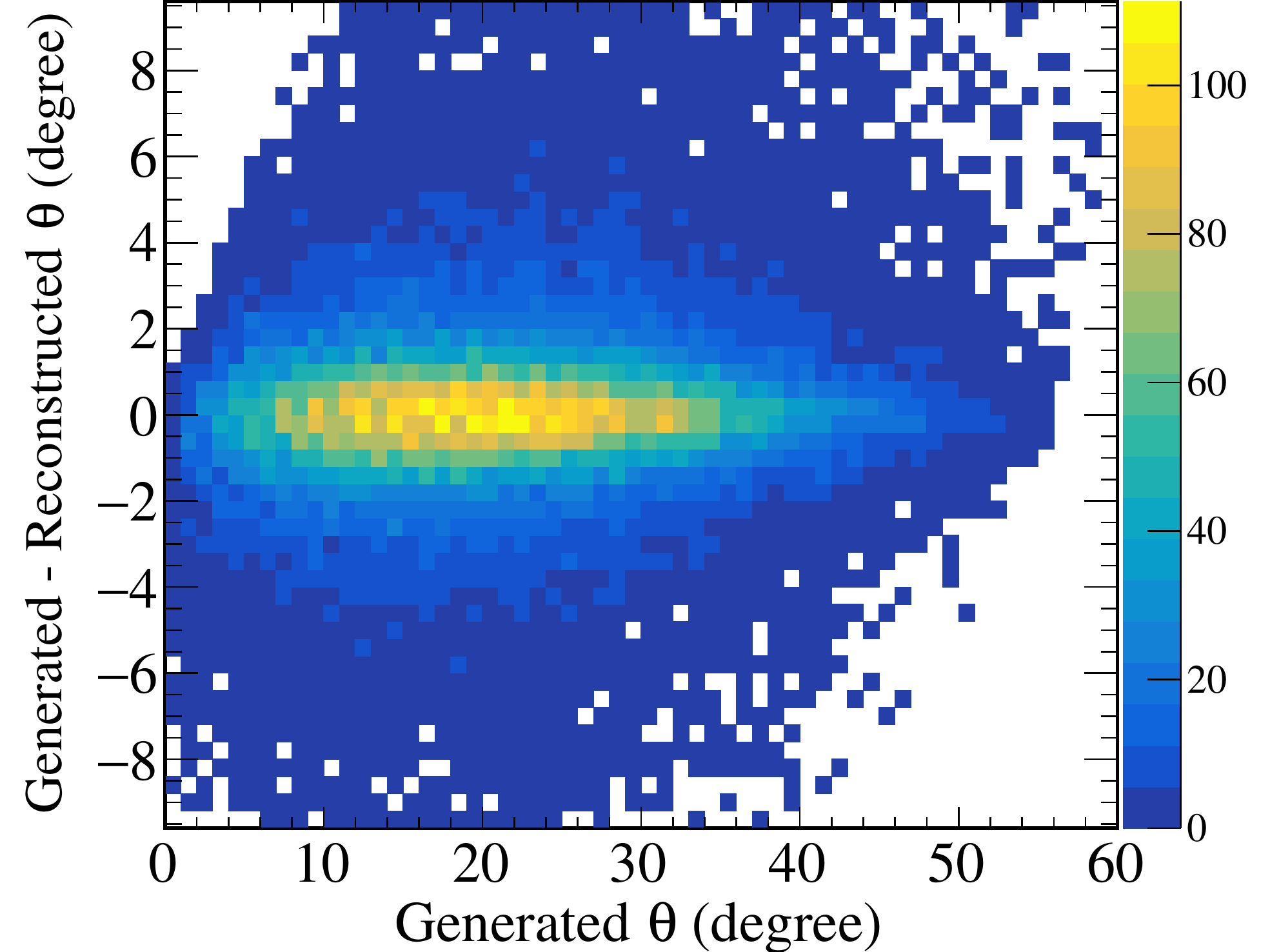}
\caption{EEE telescope angular resolution as a function of the track polar angle for different muon energy ranges and surrounding conditions. 
Top: $E_{\mu}=$(10-100 GeV), outdoors. 
Middle: $E_{\mu}=$(0.2-100) GeV, outdoors. 
Bottom: $E_{\mu}=$(0.2-100) GeV, indoors (concrete box, 30~cm walls and 70~cm roof). 
\label{resall2} } 
\end{figure}
\begin{figure}[h!]
\centering 
\includegraphics[width=1.0\columnwidth]{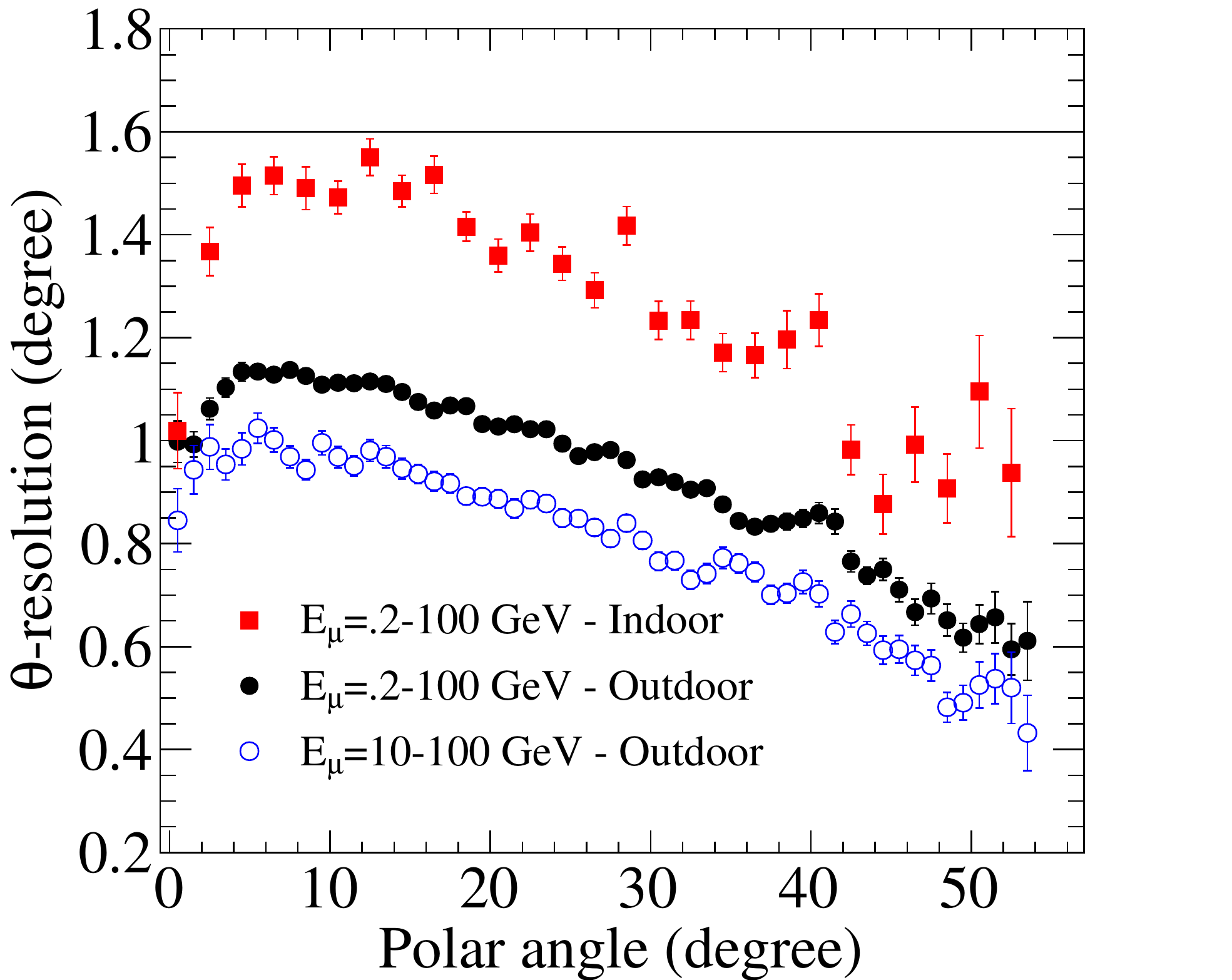}
\caption{EEE telescope angular resolution as a function of the track polar angle for different muon energy ranges and location conditions (same as in Fig~\ref{resall2}).
\label{resall2-bis} } 
\end{figure}

\begin{figure}[h!]
\centering 
\includegraphics[width=.85\columnwidth]{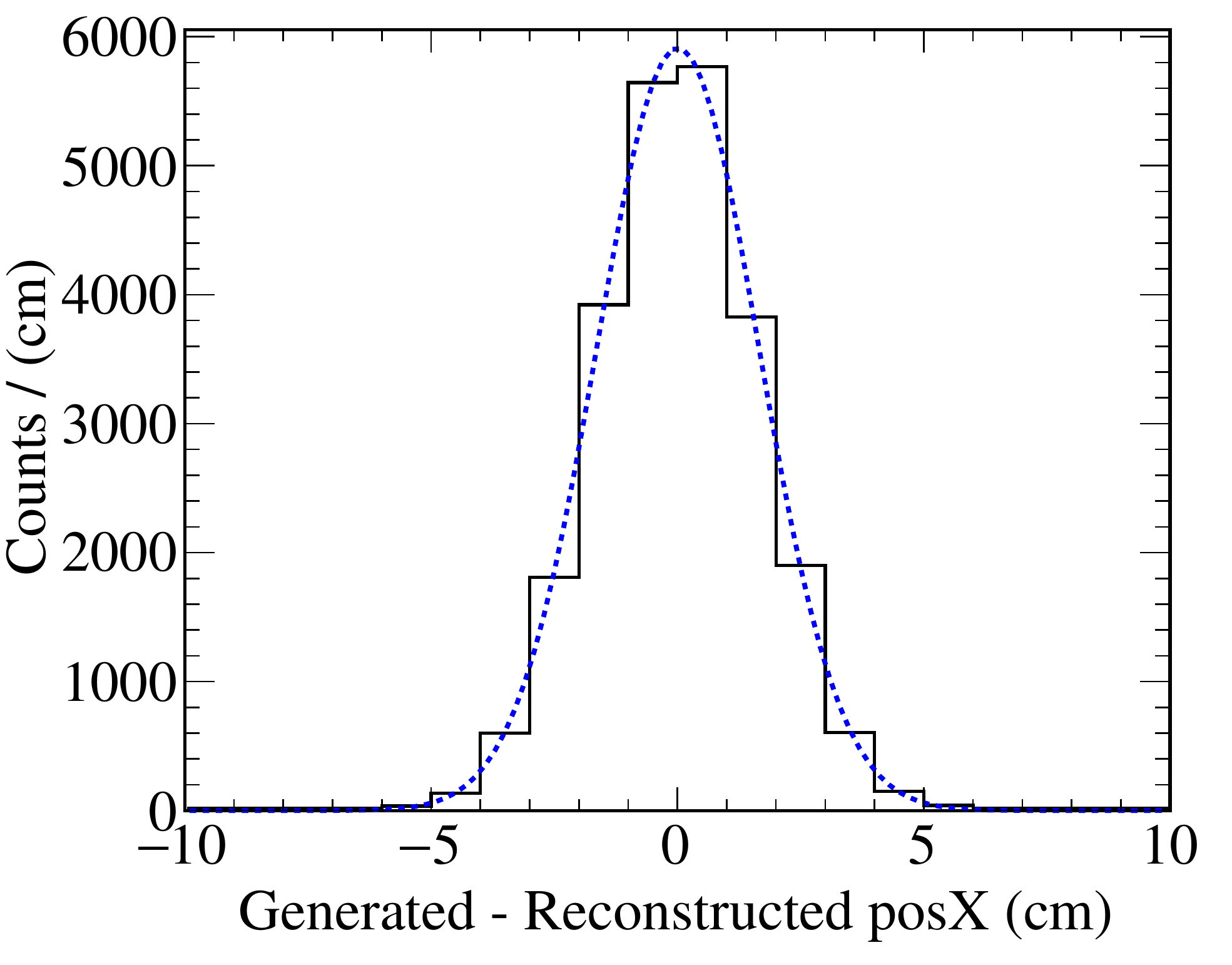} 
\includegraphics[width=.85\columnwidth]{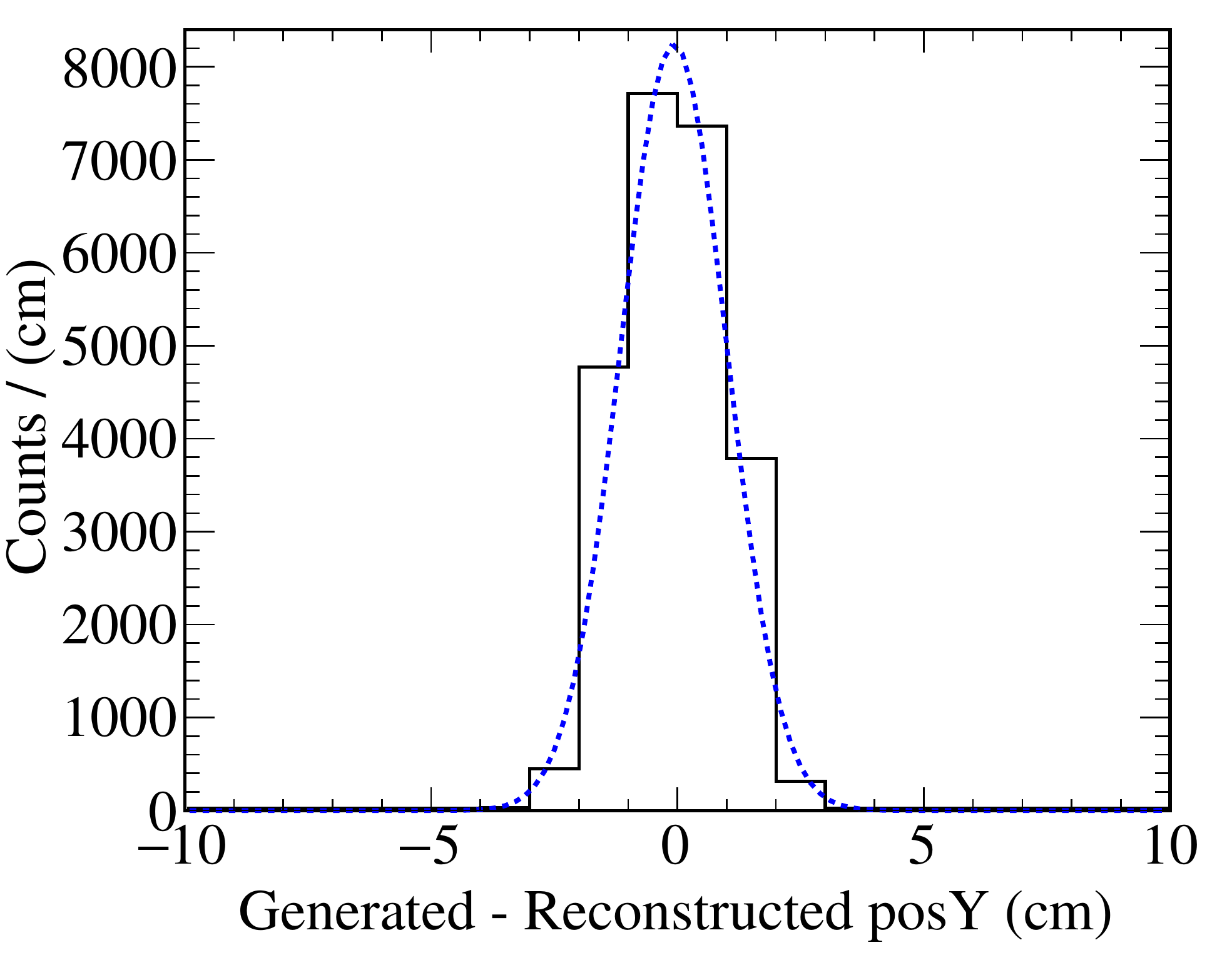} 
\caption{Spatial X (top) and Y (bottom) resolution of the middle chamber of an indoor telescope (30~cm walls and 70~cm roof, concrete) as obtained by MC simulations. Muons were generated in the (10-100) GeV energy range. The fit with a Gaussian function is shown as a dashed line. 
%The resulting parameters are $\sigma_{X} = (1.6 \pm 0.1$)~cm and $\sigma_{Y} = (1.0 \pm 0.1$)~cm, respectively. 
\label{reshig2} } 
\end{figure} 

The telescope polar angular resolution has been evaluated by generating muons according to the improved Geisser energy distribution reported in Eq.~(1), in the (10-100) GeV energy range to minimize multiple scattering of crossing particles. The telescope was simulated in the standard configuration (50/50). For the first simulation, the telescope was considered outdoors with no surrounding material. As already mentioned this is not a realistic condition since all EEE telescopes are located in rooms with roof and walls, but it is useful to assess the ultimate performance of an EEE station. The result is reported 
in Fig.~\ref{reshig1}. The telescope angular resolution, averaged over all tracks, defined as the difference between generated and reconstructed angles is $\sigma=(0.9\pm0.1)^o$. 
The dependence on the muon polar angle for different experimental setups is reported in Fig.~\ref{resall2}. The top
panel shows the best case scenario (outdoor, $E_{\mu}=$ (10-100) GeV). The middle panel shows an outdoor telescope response to a larger muon energy range, which includes the lower part of the spectrum ($E_{\mu}=$ (0.2-100) GeV). The bottom panel shows the same as above but for an indoor telescope located inside a concrete box with 30~cm -thick walls and 70~cm -thick roof (to mimic a typical room overlooked by several floors).
The angular resolution $\sigma$ as a function of the track polar angle $\theta$ is shown in Fig.~\ref{resall2-bis}. 
As expected, low energy muons undergo multiple scattering, whose effect deteriorates the performance and worsens the angular resolution by up to 50$\%$.  The best values are found at the edges of the angular acceptance. This behaviour is a non-trivial result obtained by means of the simulation framework. From this study we conclude that, even considering an indoor telescope located in a room overlooked by some concrete floors, the polar angular resolution of reconstructed tracks is always better than 1.6$^{\circ}$.

\begin{figure}[h]
\centering
\includegraphics[width=0.85\columnwidth]{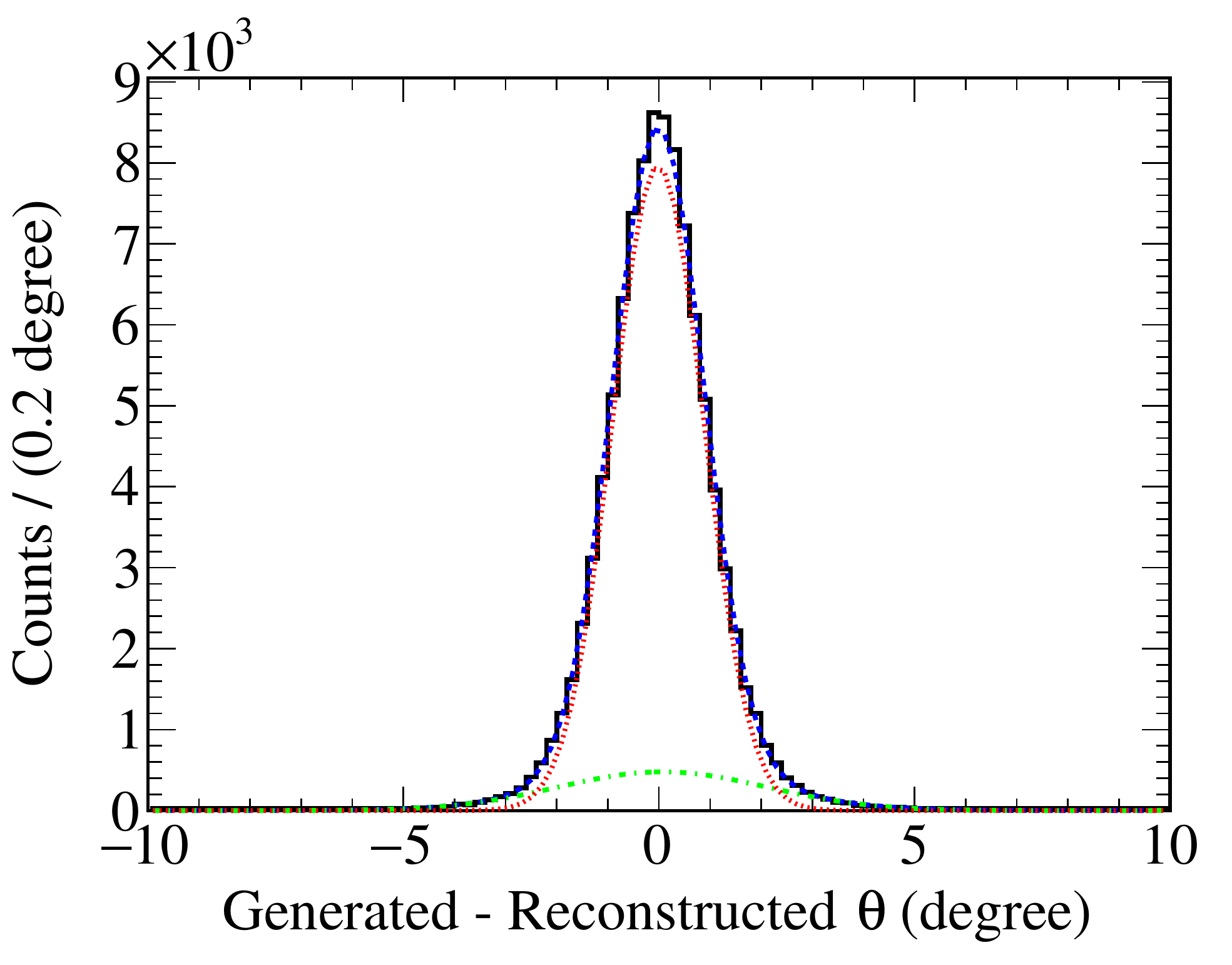}\\
\includegraphics[width=0.85\columnwidth]{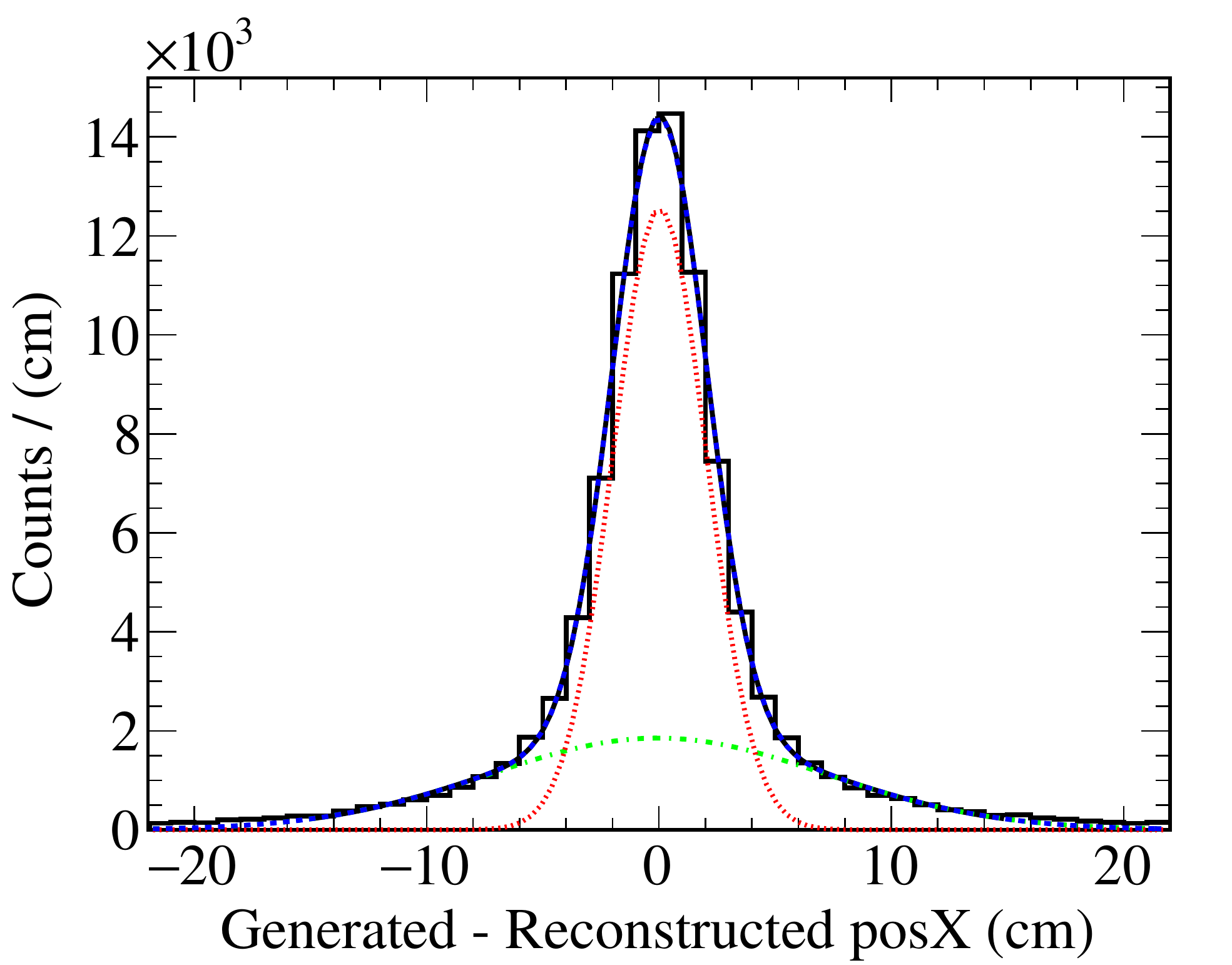}\\
\includegraphics[width=0.85\columnwidth]{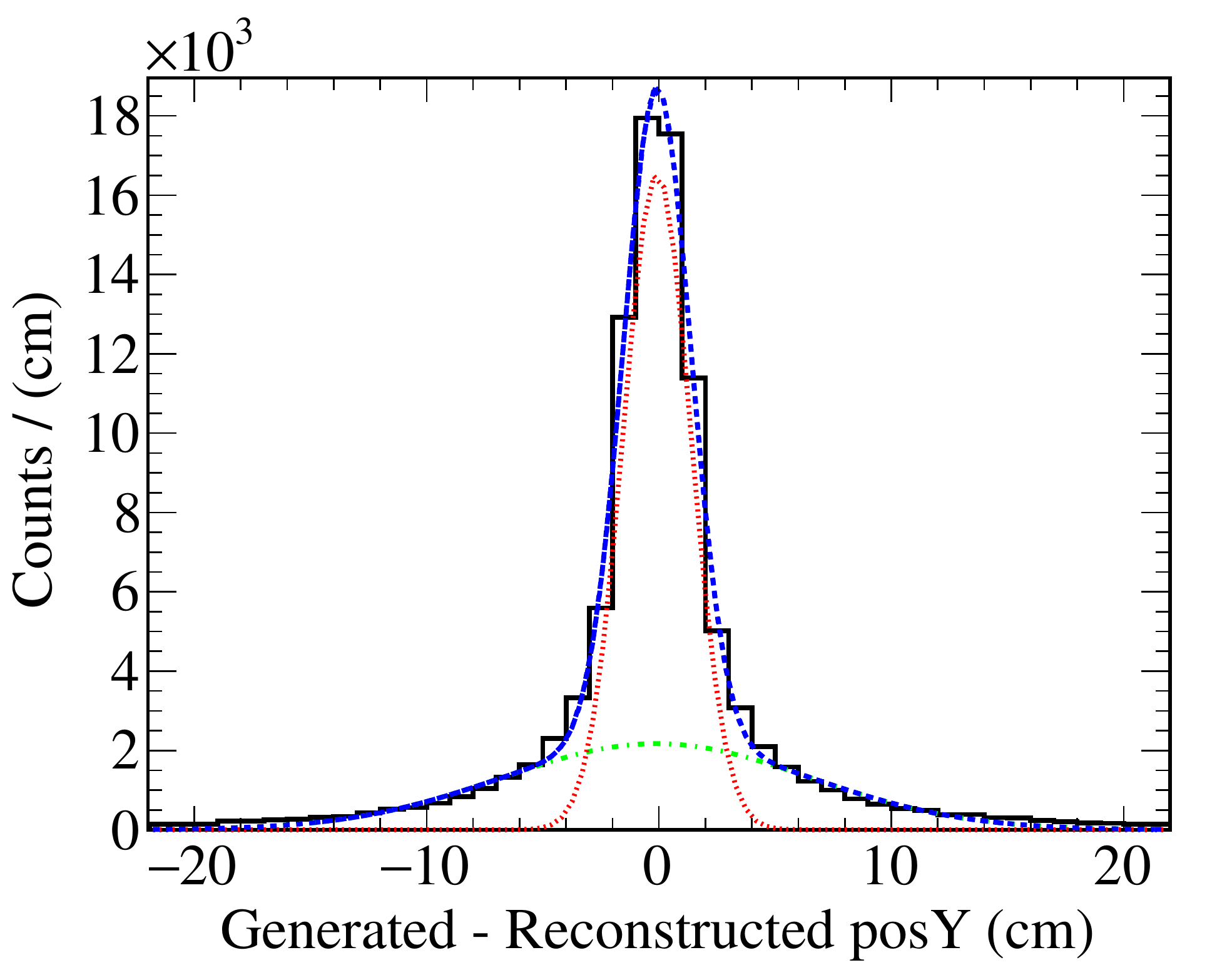}
\caption{Angular ($\theta$) and spatial (X and Y) resolutions for an outdoor telescope and muons generated in the energy range $E_\mu$ = (0.2-100) GeV. The distributions are fitted with two Gaussian lines. 
} \label{ris2}
\end{figure}

With the same procedure we extracted 
the X and Y spatial resolution for an indoor telescope (concrete box with 30~cm walls and 70~cm roof) and impinging muons in the energy range (10-100) GeV. Figure~\ref{reshig2} shows the resolution obtained for the middle chamber: $\sigma_X= (1.6\pm 0.1)$~cm and $\sigma_{Y}=(1.0\pm 0.1)$~cm.
A study performed on data (see~\cite{performance}) provided spatial resolutions $\sigma_{X}=(1.47 \pm 0.23)$~cm and $\sigma_{Y}=(0.92 \pm 0.02)$~cm that compare well to simulations.

\begin{figure}[h]
\centering
\includegraphics[width=0.85\columnwidth]{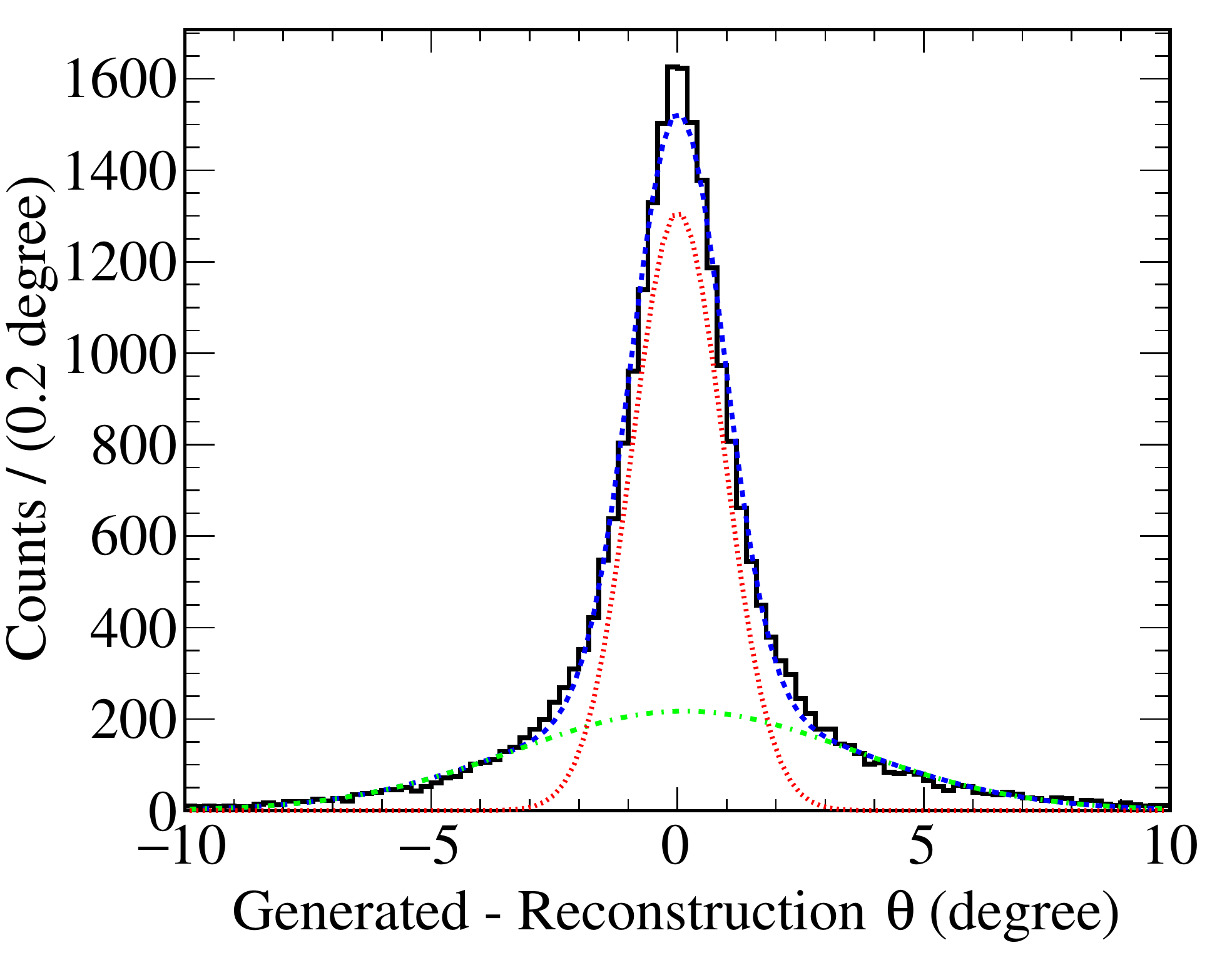}\\
\includegraphics[width=0.85\columnwidth]{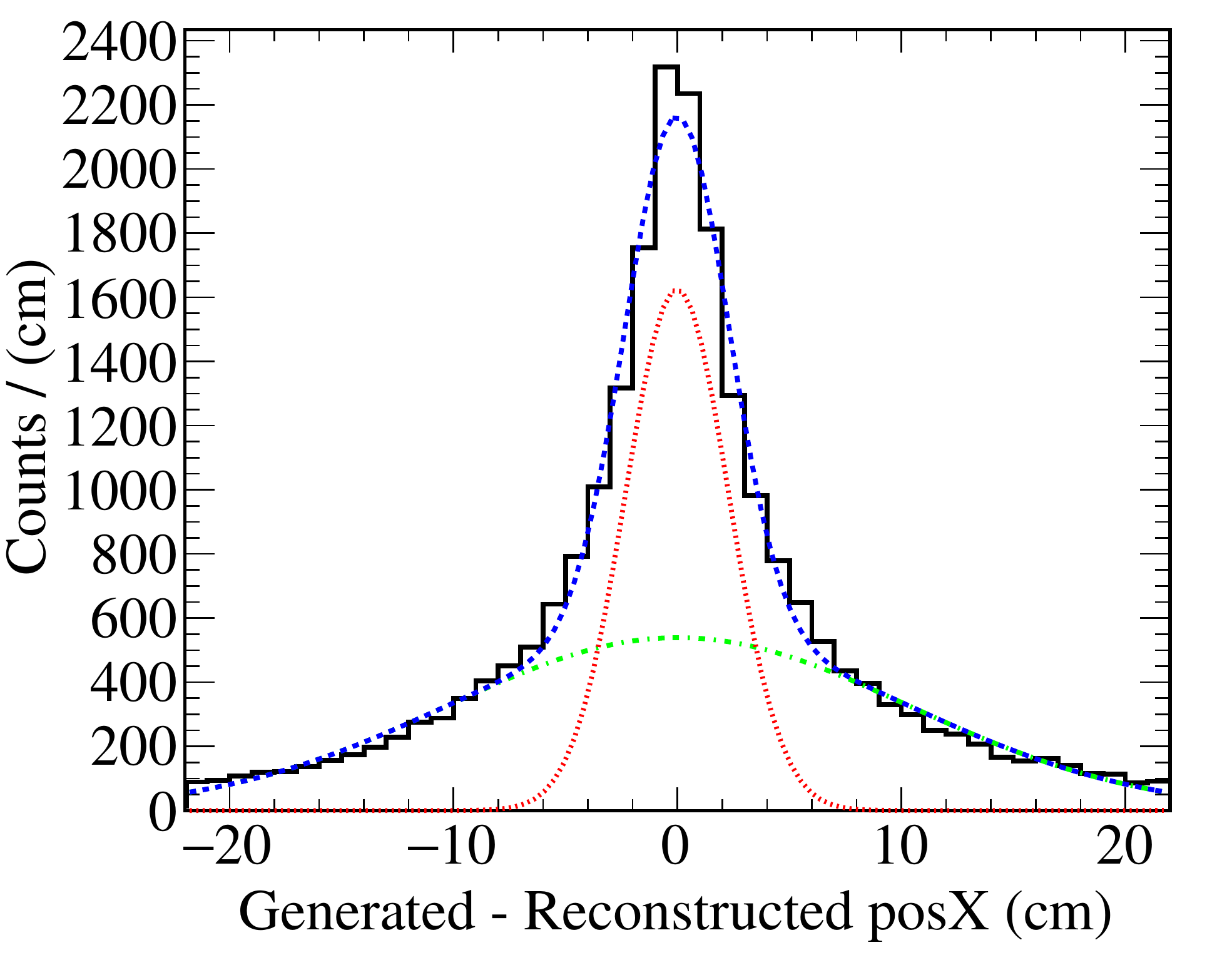}\\
\includegraphics[width=0.85\columnwidth]{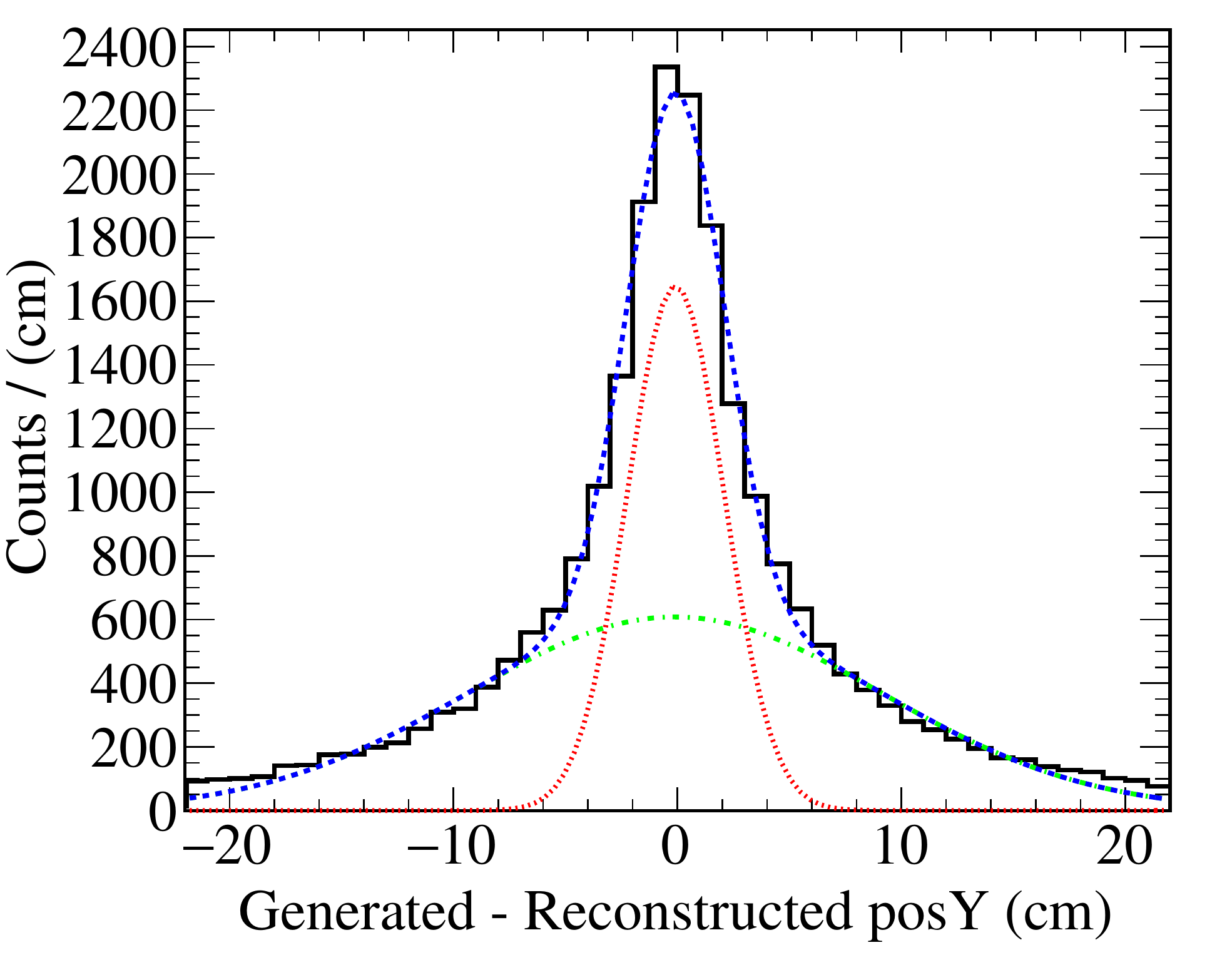}
\caption{Angular ($\theta$) and spatial (X and Y) resolution for an indoor telescope (30~cm walls and 70~cm roof, concrete) and muons generated in the energy range $E_\mu$ = (0.2-100) GeV. The distributions are fitted with two Gaussian lines.
} \label{ris3}
\end{figure}

The angular (top panel), X (middle panel) and Y (bottom panel) resolutions for an outdoor telescope and muons generated in the range (0.2-100) GeV, are shown in Fig.~\ref{ris2}. The distributions are fitted with two Gaussian functions. The two components mainly correspond to the high and low energy part of the muon spectrum.
For this experimental setup, the telescope resolutions are reported in Tab.~\ref{tab:my_label2}.
 The same study 
has been repeated for an indoor telescope (30~cm walls and a 70~cm roof, concrete) and the corresponding results are reported in Fig.~\ref{ris3} and Tab. \ref{tab:my_label2}.
As expected, spatial resolution deteriorates and both the narrow and the large component of the fit become larger.
%Table~\ref{tab:my_label2} summarises the spatial resolution for different setups.
\begin{table}[!ht]
\caption{ Angular ($\theta$) and spatial (X and Y) resolutions for different experimental conditions. Data are the standard deviations of the Gaussian fit shown in figures \ref{reshig1}, \ref{reshig2}, ~\ref{ris2} and ~\ref{ris3}.
\label{tab:my_label2}}
\centering
\begin{tabular}{c|c|c|c}
Location  & \multicolumn{2}{c|}{Outdoor} & Indoor\\ 
\hline
Muon energy (GeV)  & 10-100 & \multicolumn{2}{c}{0.2-100}\\ 
\hline\noalign{\smallskip}
    %\hline
$\theta$ (in $^o$) narrow  &   $0.9 \pm 0.1$ & $0.9 \pm 0.1$  & $0.9 \pm 0.1 $\\  
$\theta$ (in $^o$) large  & --&  $2.0 \pm 0.2 $ & $3.4 \pm 0.1 $\\  
\hline
PosX (cm)  narrow&  $1.6 \pm 0.1$ & $2.0 \pm 0.1$  & $2.3 \pm 0.1$\\  
PosX (cm) large & ---&  $7.2 \pm 0.1$ &$10.3 \pm 0.8$\\
\hline
PosY (cm) narrow & $1.0 \pm 0.1$& $1.4 \pm 0.3$ & $2.1 \pm 0.3$\\  
PosY (cm) large & --& $6.6 \pm 0.4$\ & $9.2 \pm 0.5$\\  
         % \hline
\hline\noalign{\smallskip}
\end{tabular}
\end{table}

\begin{figure}[h!]
\centering
\includegraphics[width=1.0\columnwidth]{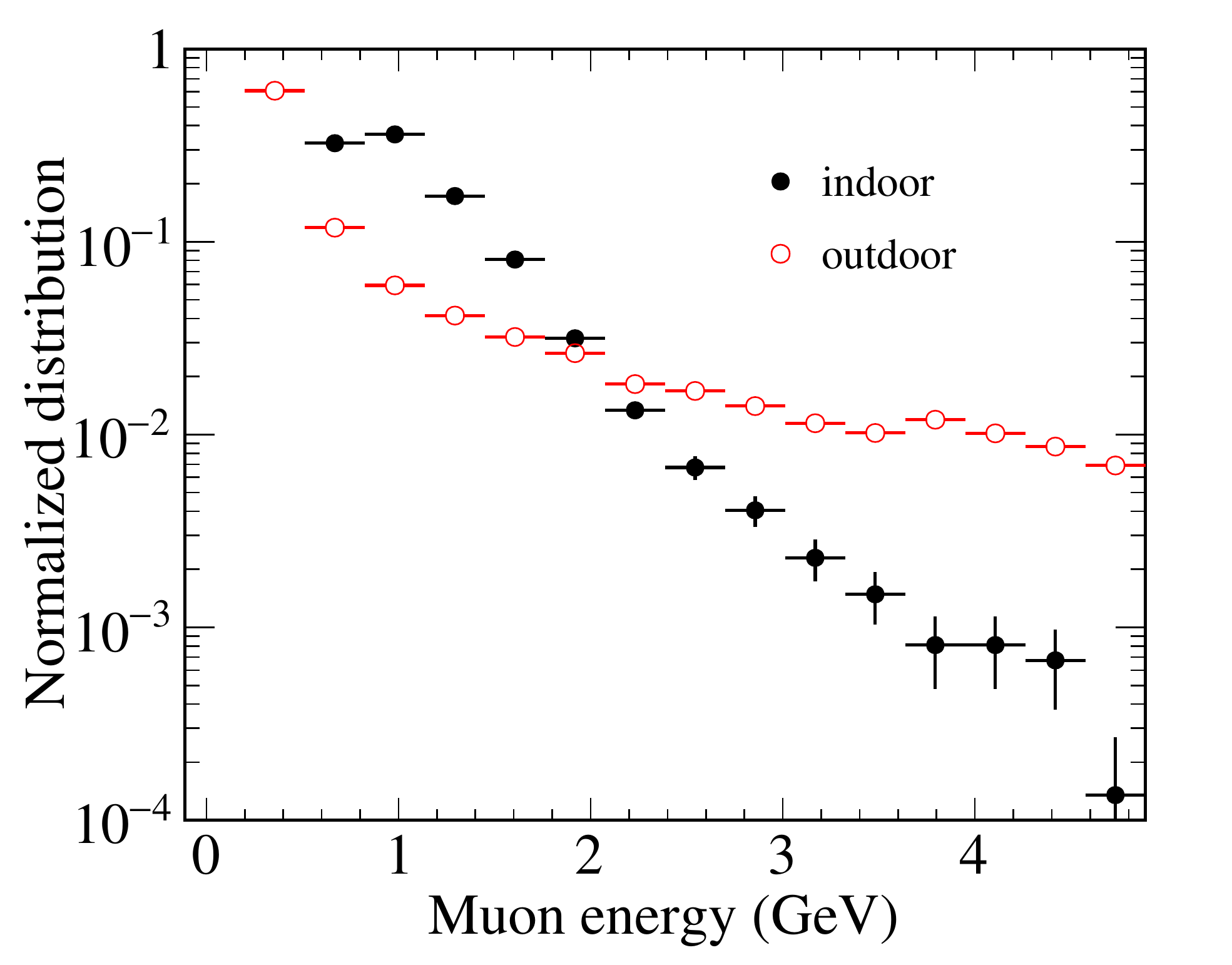}
\caption{Distributions normalized to the unit, as a function of the muon energy for muons corresponding to the large component of the double-Gaussian fit shown in Fig.~\ref{ris2} and ~\ref{ris3}. \label{in-out-res}}
\end{figure}

The effect can be better understood by plotting the energy distributions of muons that produce the large component of the fit, selecting events outside 3$\sigma$ of narrow Gaussian. Figure~\ref{in-out-res} shows the energy distributions of generated muons, normalized to the unit, for the outdoor and the indoor telescope. 
The indoor configuration shows a sharp cut at $E_{\mu}\sim0.8$ GeV due to the concrete filtering and a steep downfall that effectively limits the energy of contributing muons to $E_{\mu}<2$ GeV. The outdoor configuration shows a rapid decrease from the minimum cut-off energy in atmosphere (0.2 GeV) and a smoother behaviour that extends to larger values. 
From this comparison we conclude that the experimental resolution extracted from data depends significantly on the conditions of the measurement and in particular on the material surrounding the telescope.

%\begin{table}[!ht]
%\caption{ Angular ($\theta$) and spatial (X and Y) resolutions for different experimental conditions. A: outdoor telescope and generated muon energy 10 -100 GeV; B and D: outdoor telescope and generated muon energy (0.2-100) GeV; C and E: indoor telescope (30~cm walls and 70~cm roof, concrete) and generated muon energy (0.2-100) GeV . For B and C only the narrow component of the Gaussian fit is reported, while for D and E only the larger one.
%\label{tab:my_label2_old}}
%\centering
%\begin{tabular}{c|c|c|c}
%\hline\noalign{\smallskip}
    %\hline
%Data  & $\theta$ (in $^o$)& PosX (cm) & PosY (cm)\\  
         %\hline
%A & $0.9 \pm 0.1$ & $1.6 \pm 0.1$ & $1.0 \pm 0.1$\\   
%B & $0.9 \pm 0.1$ & $2.0 \pm 0.1$  &  $1.4 \pm 0.3$\\  
%C & $0.9 \pm 0.1 $ & $2.3 \pm 0.1$  & $2.1 \pm 0.3$\\
%D & $2.0 \pm 0.2 $ & $7.2 \pm 0.1$  & $6.6 \pm 0.4$\\ 
%E & $3.4 \pm 0.1 $ & $10.3 \pm 0.8$  & $9.2 \pm 0.5$\\ 
%         % \hline
%\hline\noalign{\smallskip}
%\end{tabular}
%\end{table}

\subsection{Efficiency}

\begin{figure}[h!]
\centering
\includegraphics[width=0.9\columnwidth]{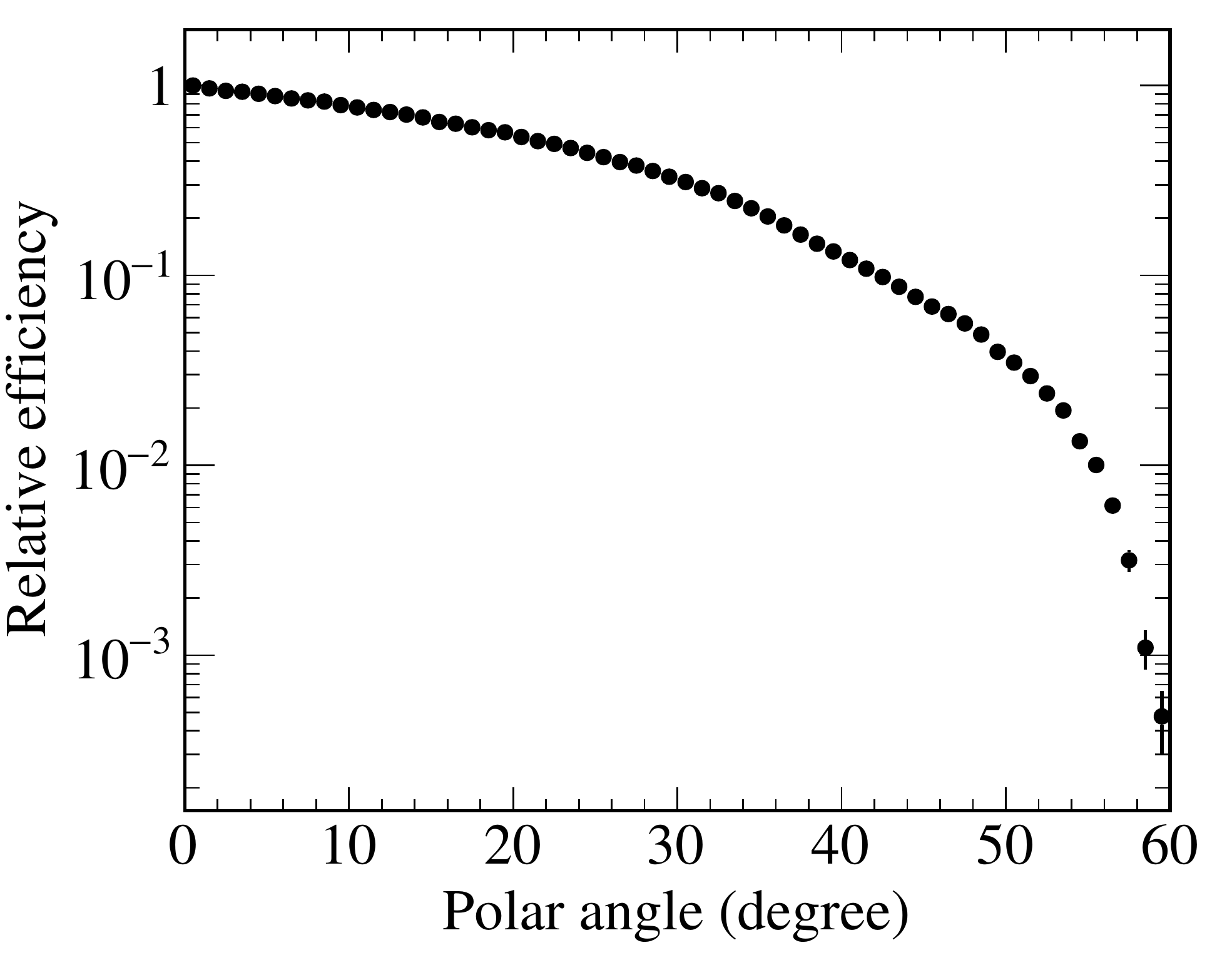}
\caption{Detection efficiency as a function of the muon track polar angle for a standard configuration of an outdoor telescope. \label{effy1}}
\end{figure}

The relative detection efficiency, obtained as the ratio of detected and generated muons, as a function of the polar angle for an outdoor standard configuration (50/50~cm) telescope, and muons generated in the full energy range (0.2-100 GeV) is shown in Fig.~\ref{effy1}. The distribution has been normalized to the maximum value corresponding to vertical tracks. The relative detection efficiency monotonically decreases as the track polar angle increases.

\begin{figure}[h!]
\centering
\includegraphics[width=0.9\columnwidth]{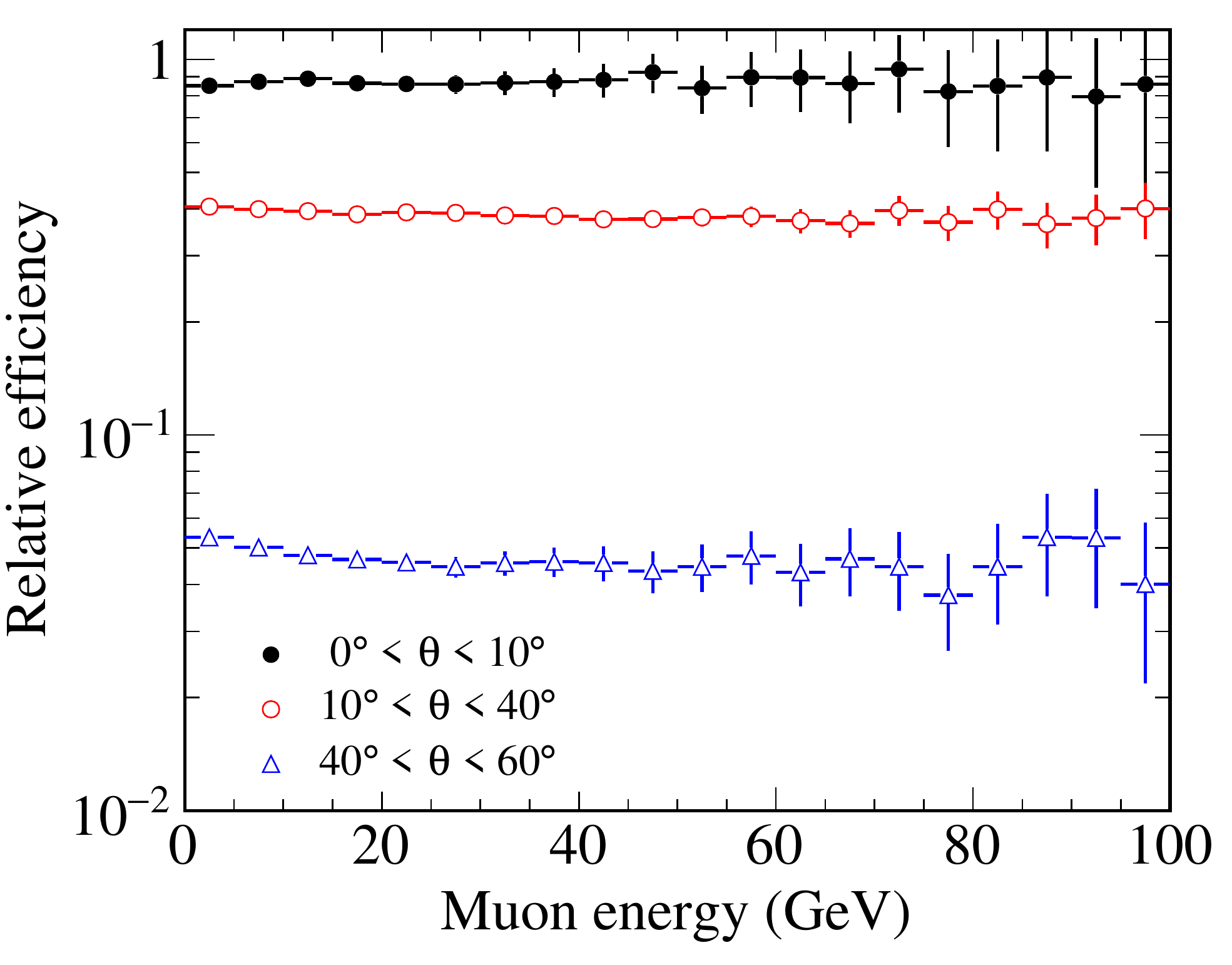}
\caption{Detection efficiency as a function of muon energy for different intervals of the track polar angle for a standard configuration of an outdoor telescope (see the insert for symbols explanation).\label{effy2}}
\end{figure}

The relative detection efficiency
as a function of muon energy for different intervals of track polar angle is shown in Fig.~\ref{effy2}. The distributions are normalized to the maximum value corresponding to $\theta=0^o$.
The relative detection efficiency shows a weak dependence on the muon energy for all angular intervals.

\subsection{Effects of surrounding materials}
\label{env_condition}

\begin{figure}[h!]
\centering
\includegraphics[width=0.85\columnwidth]{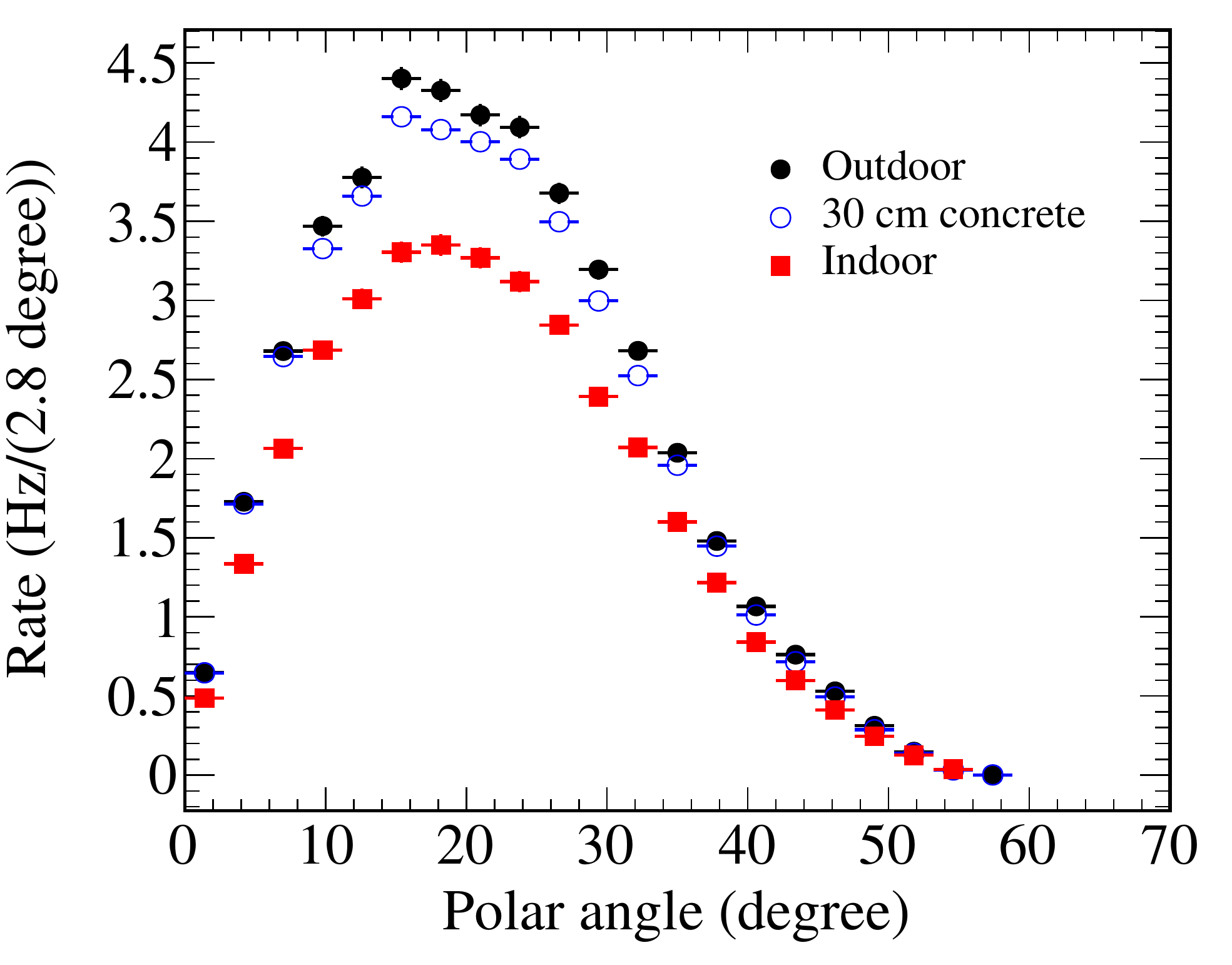}
\caption{Comparison of absolute single muon rates as a function of tracks polar angle in different simulation environmental conditions. \label{in-outdoor}}
\end{figure}

As already mentioned, the EEE telescopes are often placed in rooms with variable thickness of concrete walls and roof. Moreover, detailed drawings of the building (schools, university labs, ...) are not always available for a thorough assessment of the experimental conditions. For a correct comparison of different telescopes it is therefore important to evaluate the effect of the location.

We studied different experimental setups, common to the EEE network, with an increasing complexity of the environmental conditions. We started by comparing the results obtained simulating an outdoor and an indoor (30~cm walls and a 70~cm roof, concrete) telescope.
The simulated single muon absolute rates as a function of tracks polar angle are shown in Fig.~\ref{in-outdoor}. The indoor configuration shows a significant absorption for tracks in the range (10-30)$^o$. As a cross check, results from an intermediate configuration (30~cm of concrete for both walls and roof) are also reported. To have an idea of the overall effect, the predicted single muon rates at sea level are: 45.2 Hz for an outdoor telescope, 35.1 Hz for indoor and 43.2 Hz for an intermediate configuration. 
\begin{figure}[h!]
\centering
\includegraphics[width=0.85\columnwidth]{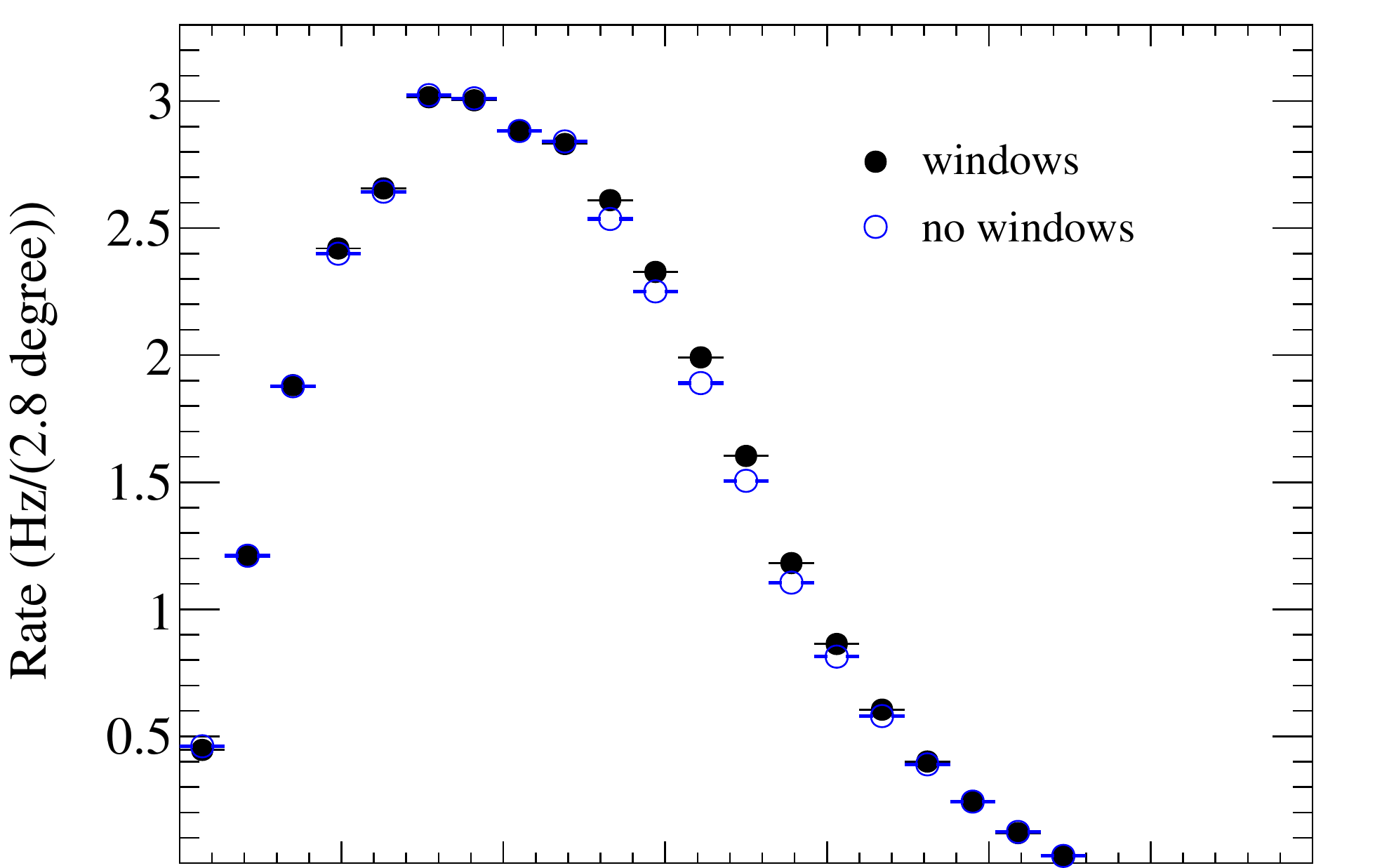}\\
\vspace{-1mm}
\includegraphics[width=0.85\columnwidth]{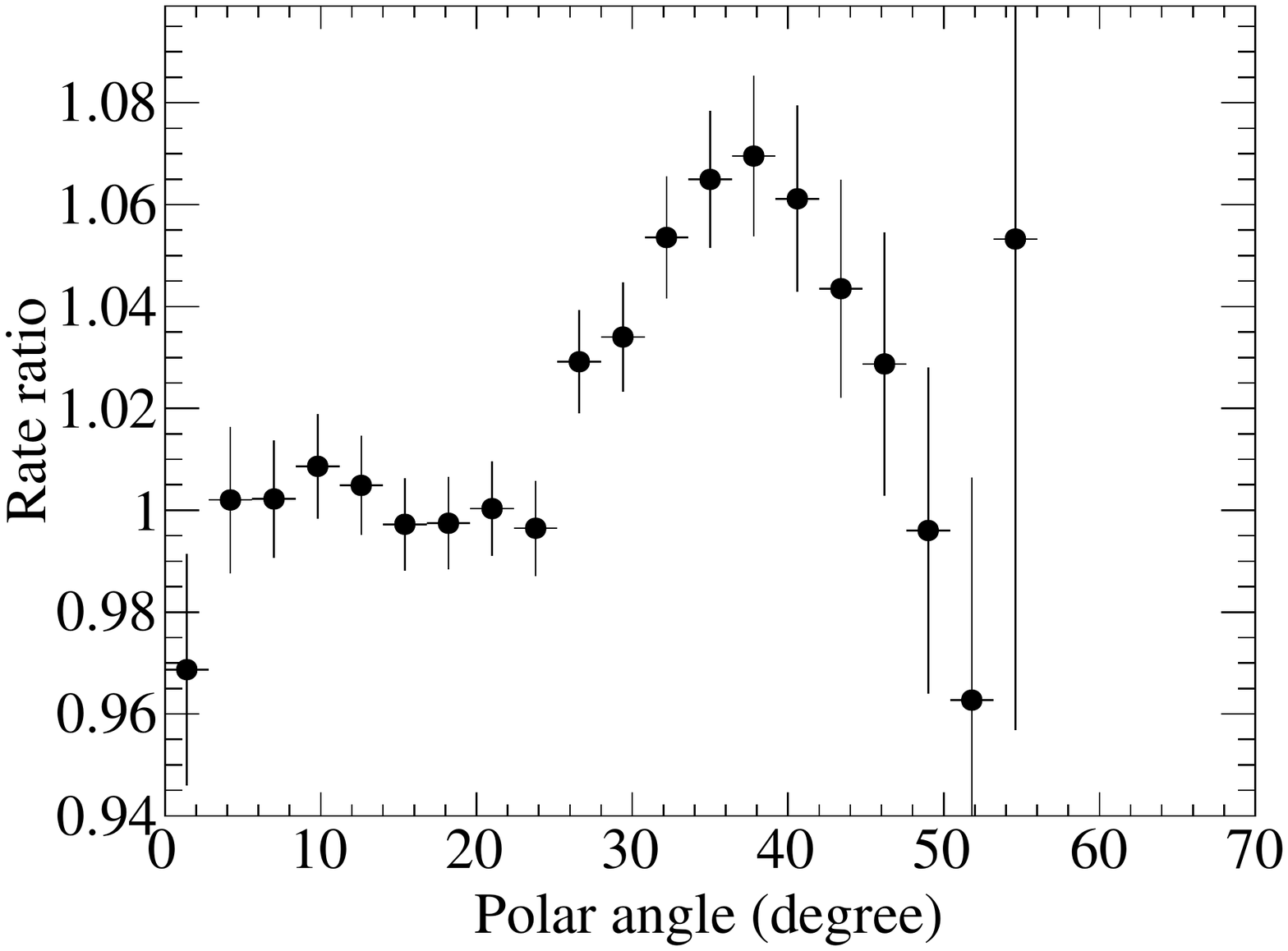}\\
\includegraphics[width=0.48\columnwidth]{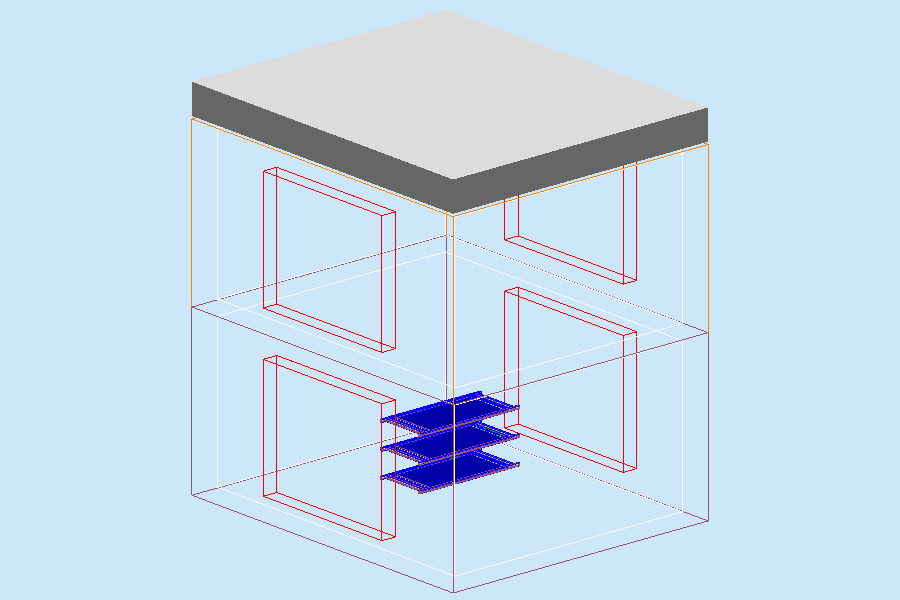}
\includegraphics[width=0.48\columnwidth]{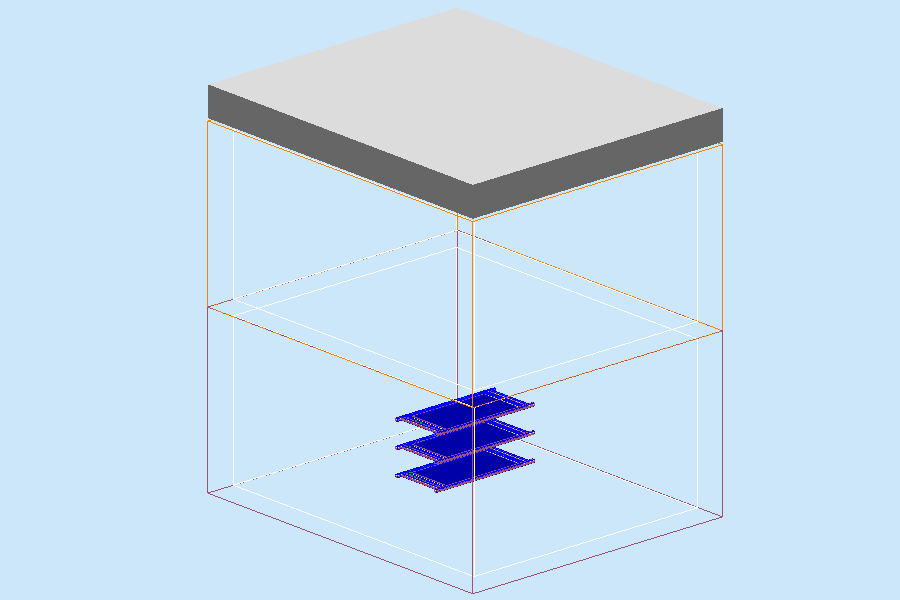}
\caption{Top, counting rate as a function of tracks polar angle for an indoor telescope hosted in a building with large windows (full circles) and without (open circles). Middle, ratio between the two rate distributions. Bottom, the two simulated geometries. \label{windows}}
\end{figure}

The comparison between an indoor telescope hosted in a room with and without windows is shown in Fig.~\ref{windows}. 
In the top panel of the figure, the simulated single muon rates are shown as a function of the track polar angle, in the middle the ratio between the two, and in the bottom panel a picture with the two simulated geometries is shown, with the windows positions in evidence.
Simulation results show that even such a little detail affects
the measured angular distributions, with a non trivial distortion that needs to be accounted for in order to keep the systematic error on experimental data as low as possible. Such a significant sensitivity, is an interesting feature that demonstrates how muons can be used to perform a scan of surrounding materials ({\it muon tomography}) ~\cite{Riggi}. Further investigations are in progress and results will be present in a dedicated paper.

\begin{figure}[h!]
\centering
\includegraphics[width=1.0\columnwidth]{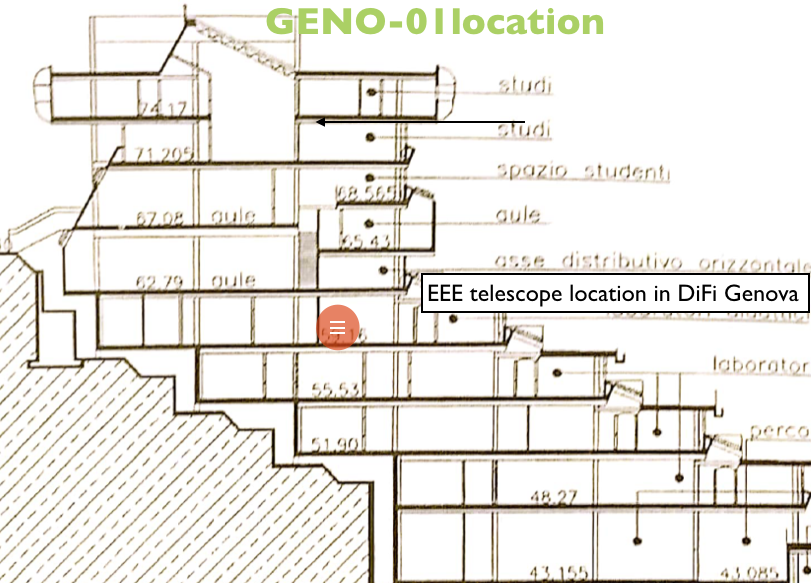}\\
\caption{A cross section of the Physics Department of University of Genoa with, in evidence, the location of the EEE telescope therein. 
 \label{gen}}
\end{figure}

To further investigate these effects in a more realistic case, we considered the telescope installed at the Physics Department of the University of Genova (GENO-01).
This is a particular complicated geometry, since the telescope is located under many floors and the building is surrounded by a mountain on one side and a valley on the 
other, defining an asymmetric shielding for cosmic rays. A drawing of the building is reported in Fig.~\ref{gen}. 

\begin{figure}[h!]
\centering
\includegraphics[width=0.9\columnwidth]{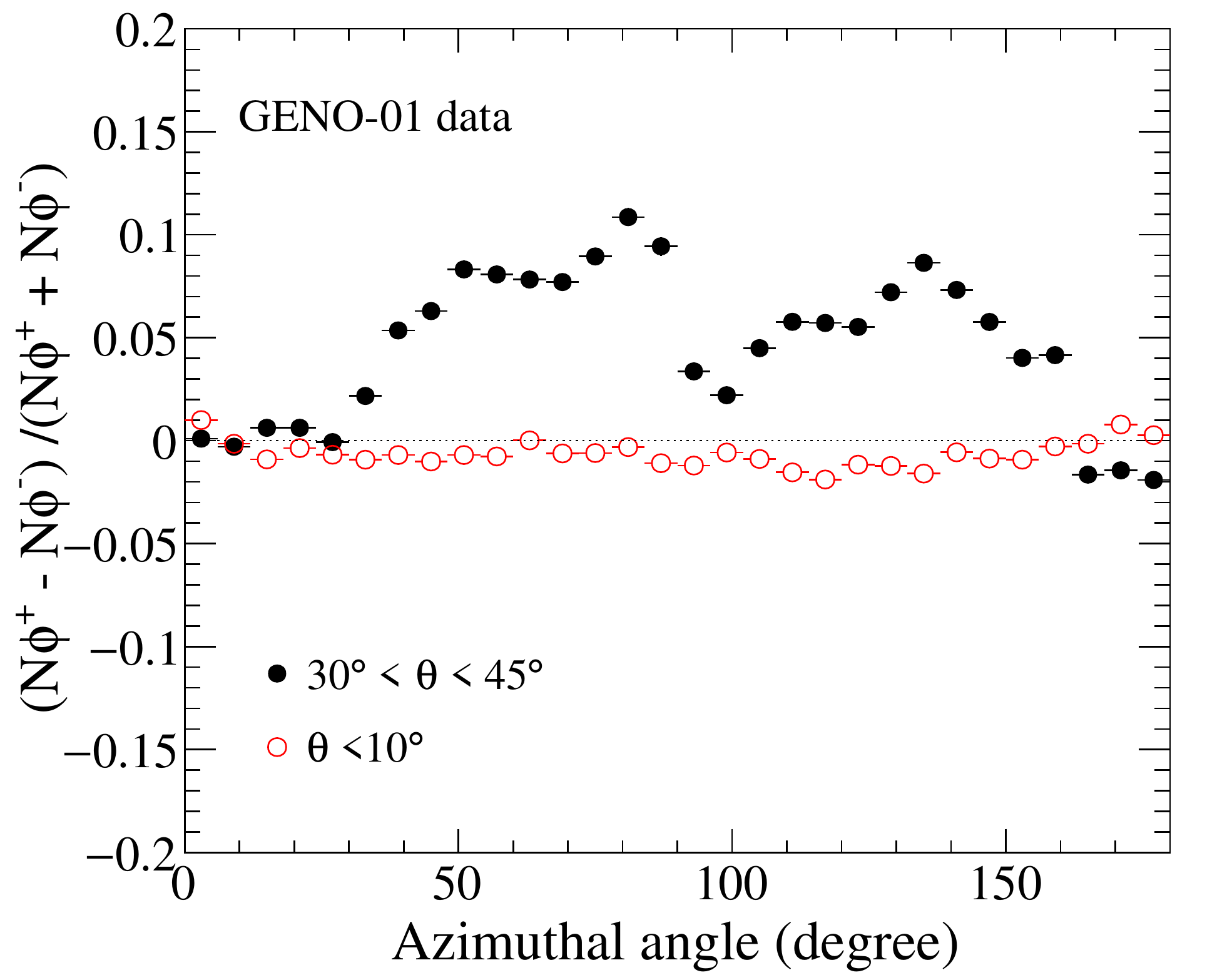}
\caption{GENO-01 telescope experimental asymmetry for muon tracks with $30^{\circ}<\theta< 45^{\circ}$ (full circles), with $\theta< 10^{\circ}$ (open circles). The azimuthal angle is defined such as $\phi=0^o$ corresponds to the detector X-axis. N$_{\phi^+}$ is the yield for tracks from the valley direction ( $0^{\circ}<\phi\le 180^{\circ}$) while N$_{\phi^-}$- accounts for tracks from the hill side ($0^{\circ}> \phi \ge -180^{\circ}$). 
\label{gen1}}
\end{figure}

Selecting measured muon tracks in the polar angle range $30^{\circ}<\theta< 45^{\circ}$, an 
asymmetry of the order of $10\%$ is present in the azimuthal angle distribution corresponding to valley-side tracks ($0^{\circ}<\phi\le 180^{\circ}$) and hill-side tracks ($0^{\circ}> \phi \ge -180^{\circ}$). Figure~\ref{gen1} shows the experimental asymmetry as measured by the GENO-01 telescope. The asymmetric absorption of muons due to the building structure may explain  the observed asymmetry. 
As shown in Fig.~\ref{gen1} (open circles), the asymmetry disappears when almost vertical tracks ($\theta< 10^{\circ}$) are selected, as expected. In this condition, muons with any $\phi$ cross roughly the same absorber thickness providing a flat azimuthal distribution.

This configuration was implemented in the simulation framework by placing the detector below a 70~cm-thick concrete asymmetric roof 
(mimicking the asymmetric structure of the building)
and placing on one side a volume that %fully 
absorbs muons (mimicking the side hill). 
\begin{figure}[h!]
\centering
\includegraphics[width=0.68\columnwidth]{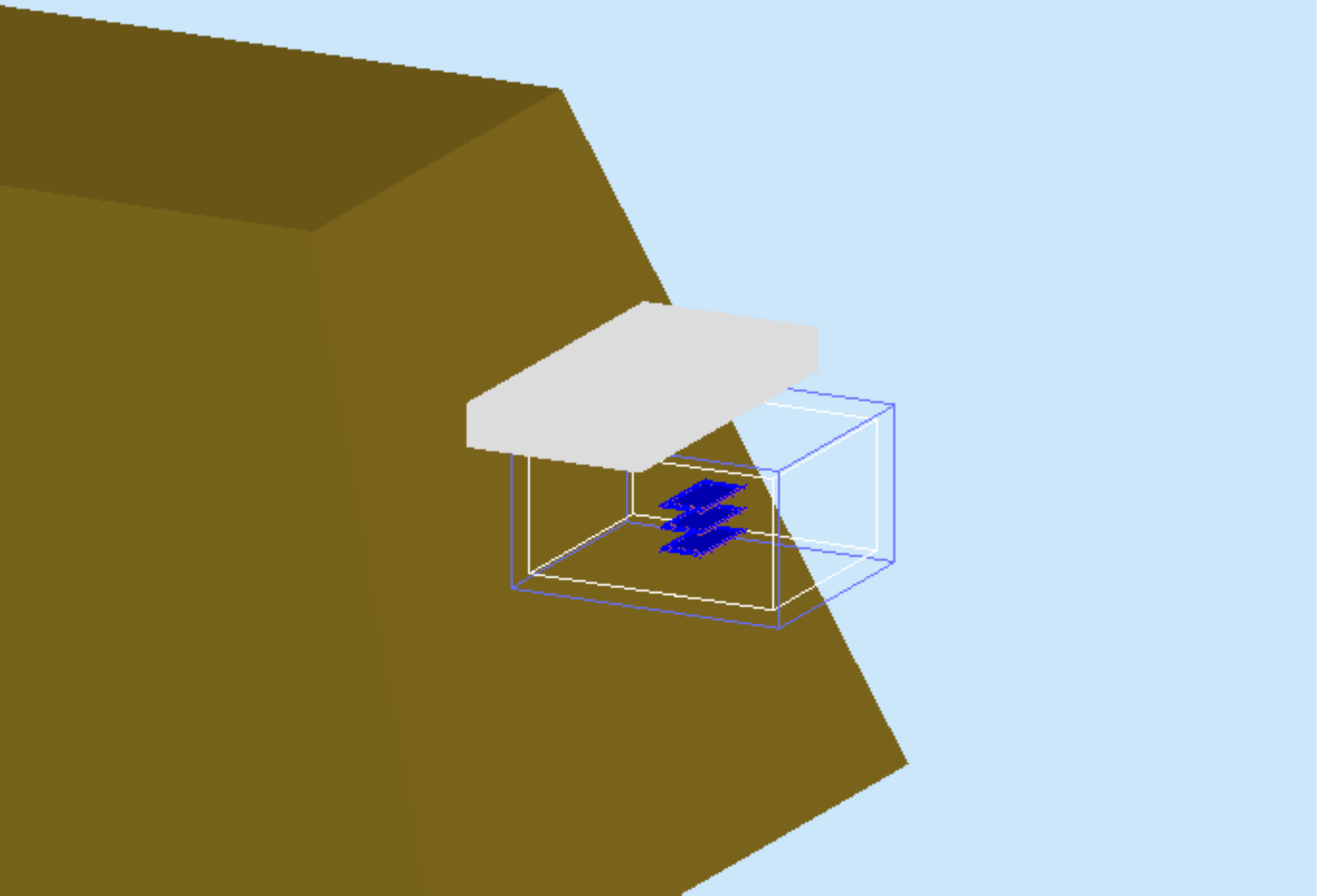}
\includegraphics[width=0.9\columnwidth]{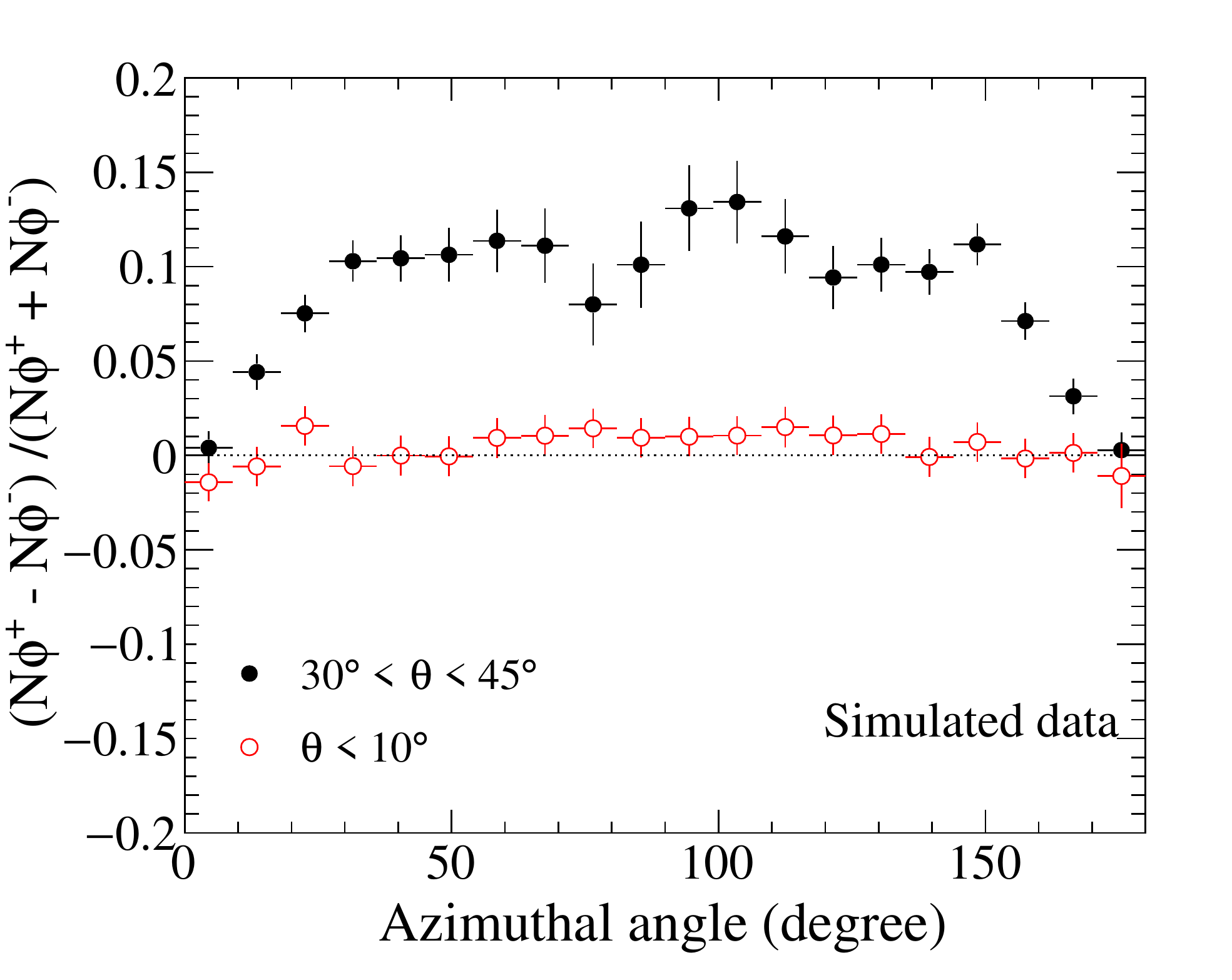}
\caption{Top panel: the building geometry implemented in the simulation framework. Bottom panel: the resulting $\phi$ asymmetry (definitions are the same as in Fig.~\ref{gen1}). 
\label{gensim}}
\end{figure}
The GEMC implementation and the resulting asymmetry are shown in Fig~\ref{gensim}. The simulations are able to reproduce shape and magnitude of the experimental asymmetries for $\theta<10^\circ$ and $30^\circ<\theta<45^\circ$.
As a cross check, we calculated the asymmetry for a standard outdoor telescope finding, as expected,that 
the asymmetry is zero for both $\theta<10^\circ$ and $30^\circ<\theta<45^\circ$ polar angle intervals, as shown in Fig.~\ref{gensim2}.
\begin{figure}[h!]
\centering
\includegraphics[width=0.85\columnwidth]{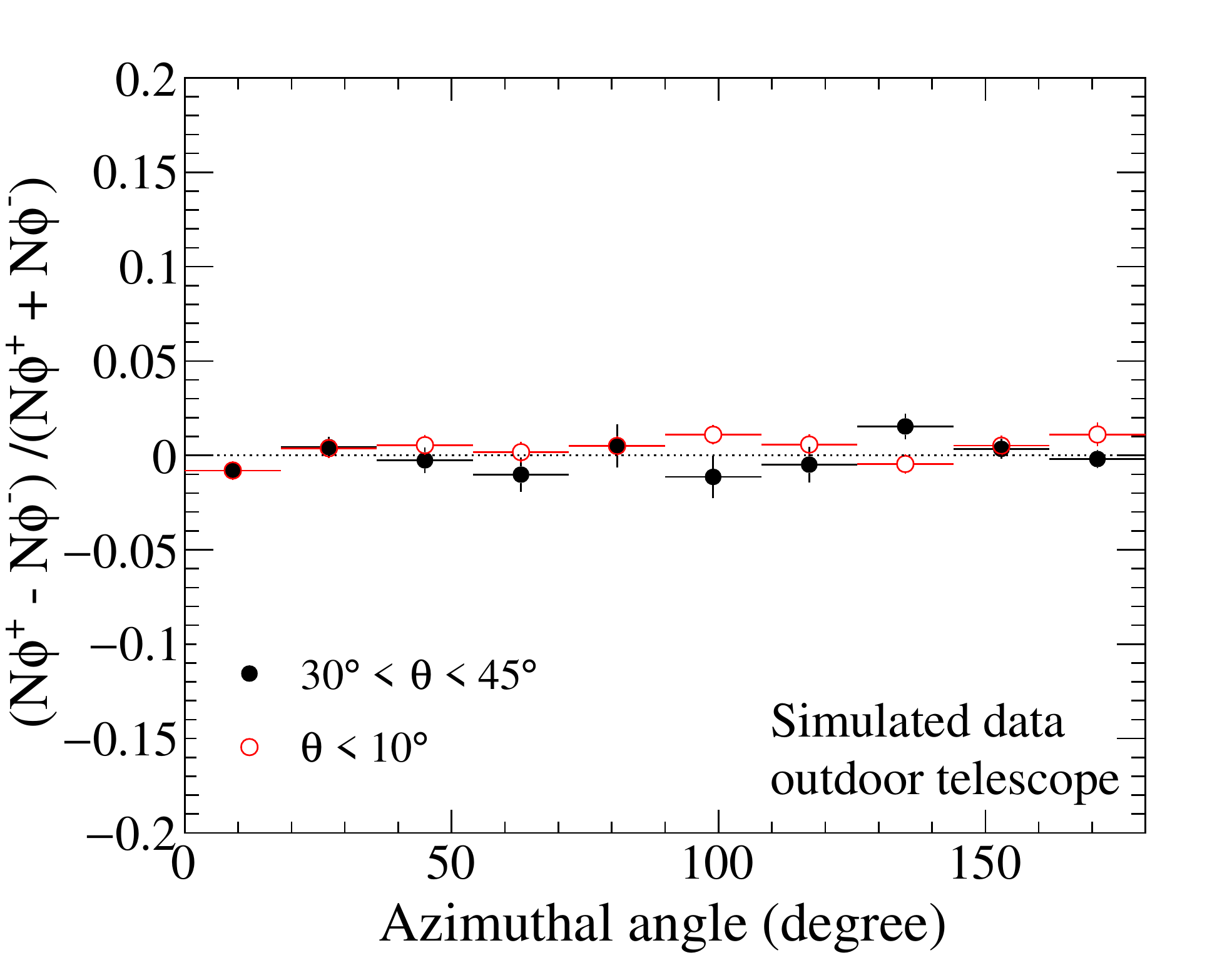}
\caption{Simulated azimuthal asymmetry for an outdoor telescope and muon tracks with $30^{\circ}<\theta< 45^{\circ}$ full circles, $\theta< 10^{\circ}$ open circles. \label{gensim2}}
\end{figure}

This study demonstrates that the EEE simulation framework is able to reproduce data collected by EEE telescopes even in complicated experimental set-ups.

\section{Conclusions}
\label{conc}
The EEE Collaboration has developed a full simulation framework to study the response of the cosmic muon telescopes of its network. MRPC geometry, materials and a parametric response of the detectors have been implemented in GEMC, a user-friendly interface to GEANT4 libraries.
The framework includes a cosmic muon event generator based on an improved Gaisser parametrization of the muon flux at the Earth level. The framework has been validated by a detailed comparison to single-muon rates (angular and integrated) recorded by some selected EEE telescopes. The agreement between experimental data and simulated data at few percent on several observables (absolute cosmic rate, angular distributions, resolution, ...), gives us confidence on the validity of our implementation.\\
The EEE simulation framework is a valuable tool to study the detector performance in terms of: efficiency, angular and spatial resolutions, and dependence on telescope set-up (including detector geometry and surrounding materials) \footnote{Simulations can be used to understand the effect of the single telescope resolution on the the whole EEE network sensitivity to cosmic rays direction.}.
Simulations can be used to compare and correct the response of different detectors in the EEE network in order to achieve the systematic precision requested by the study of small effects such as the variation of the cosmic ray flux due to the Forbush effect.
Finally, the EEE simulation framework can be used to investigate new directions, such as the use of the cosmic muons for building tomography, extending the current scope of the EEE Collaboration.

\section*{Acknowledgments}
\label{acknow}
This work was supported by the Museo Storico della Fisica e Centro Studi e Ricerche Enrico Fermi. M.B.and M.U. are supported by the U.S. Department of Energy, Office of Science, Office of Nuclear Physics under contract DE-AC05-06OR23177.

\end{document}